\documentclass[12pt,a4paper]{book}
\usepackage{makeidx}     
\usepackage{graphicx}    
\usepackage{multicol}

\makeindex

\begin{document}


\vskip 1.5cm \centerline{\bf \large ELECTRONIC STRUCTURE OF
SPHEROIDAL FULLERENES }  \vspace{15mm}


\centerline{\bf RICHARD PINCAK$^{a,b}$ and MICHAL PUDLAK$^{a}$ }
\vspace{8mm}

\centerline{$^{a}$Institute of Experimental Physics, Slovak Academy
of Sciences,} \centerline{ Watsonova 47,043 53 Kosice, Slovak
Republic} \centerline{$^{b}$Bogoliubov Laboratory of Theoretical
Physics,} \centerline{Joint Institute for Nuclear Research, Dubna,
Russia} \vspace{20mm}

\centerline{\bf \large Abstract}

Both the eigenfunctions and the local density of states (DOS) near
the pentagonal defects on the fullerenes \index{fullerenes} was
calculated analytically as a numerically. The results shows that the
low-energy DOS has a cusp which drops to zero at the Fermi energy
for any number of pentagons at the tip except three. For three
pentagons, the nonzero DOS across the Fermi level
\index{level!Fermi}is formed. Graphite is an example of a layered
material that can be bent to form fullerenes which promise important
applications in electronic nanodevices\index{nanodevices}. The
spheroidal geometry of a slightly elliptically deformed sphere was
used as a possible approach to fullerenes. We assumed that for a
small deformation the eccentricity of the spheroid is much more
smaller then one. We are interested in the elliptically deformed
fullerenes C$_{70}$ as well as in C$_{60}$ and its spherical
generalizations like big C$_{240}$ and C$_{540}$ molecules. The
low-lying electronic levels are described by the Dirac equation in
(2+1) dimensions. We show how a small deformation of spherical
geometry evokes a shift of the electronic spectra compared to the
sphere and both the electronic spectrum of spherical and the shift
of spheroidal fullerenes \index{fullerenes!spheroidal} were derived.
In the next study the expanded field-theory model was proposed to
study the electronic states near the Fermi energy in spheroidal
fullerenes. The low energy electronic wave functions obey a
two-dimensional Dirac equation on a spheroid with two kinds of gauge
fluxes taken into account. The first one is so-called K spin flux
which describes the exchange of two different Dirac spinors in the
presence of a conical singularity. The second flux (included in a
form of the Dirac monopole field) is a variant of the effective
field approximation for elastic flow due to twelve disclination
defects through the surface of a spheroid. We consider the case of a
slightly elliptically deformed sphere which allows us to apply the
perturbation scheme. We shown exactly how a small deformation of
spherical fullerenes provokes an appearance of fine structure in the
electronic energy spectrum as compared to the spherical case. In
particular, two quasi-zero modes in addition to the true zero mode
are predicted to emerge in spheroidal fullerenes. The effect of a
weak uniform magnetic field on the electronic structure of slightly
deformed fullerene molecules was also studied. It was shown how the
existing due to spheroidal deformation fine structure of the
electronic energy spectrum splitted in the presence of the magnetic
field. We shown that the fine structure of the electronic energy
spectrum is very sensitive to the orientation of the magnetic field.
We found that the magnetic field pointed in the x direction does not
influence the first electronic level whereas it causes a splitting
of the second energy level. Exact analytical solutions for
zero-energy modes are found. The HOMO (highest occupied molecular
orbital)-LUMO (lowest unoccupied molecular orbital) gap was
calculated within the continuum field-theory model of fullerenes.
\vspace{8mm}

\chapter[Electronic Structure of
Spheroidal Fullerenes]{Introduction} \label{chap:Intro}

The purpose of this chapter is to present some examples of using
topology and geometry in a study of a new interesting class of
carbon materials as fullerenes. Fullerenes are cage-like molecules
of carbon atoms. The discovery of these molecules has attracted
considerable attention of both experimentalists and theorists due to
their unique physical properties. An further interest to carbon
nanoparticles originates from the fact that the geometry is
accompanied by topological defects. The topologically nontrivial
objects play the important role in various physically interesting
systems. The 't Hooft-Polyakov monopole in the non-Abelian Higgs
model, instantons in quantum chromodynamics, solitons in the Skyrme
model, Nielsen-Olesen magnetic vortices in the Abelian Higgs
model(see, e.g.,~\cite{rajaraman}) are examples of such
topologically nontrivial objects. Notice that the topological origin
the elastic flux results in appearance of the disclination-induced
Aharanov-Bohm like phase (see~\cite{Volod}). In condensed matter
physics for instance, vortices in liquids and liquid crystals,
solitons in low-dimensional systems are objects with nontrivial
topology.

\begin{figure}[t]
\begin{center}
\includegraphics[scale=.5]{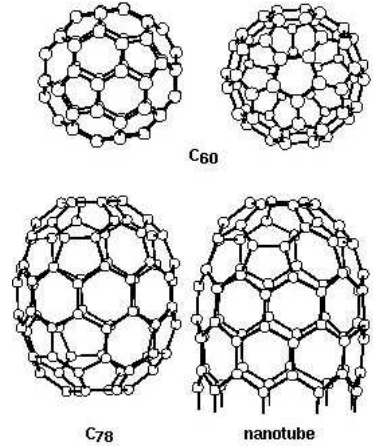}
\caption{The symmetric structure of fullerene C$_{60}$, elliptically
deformed molecule C$_{78}$ and big deformed structure as carbon
nanotube. \label{fig1}}
\end{center}
\end{figure}
\begin{figure}[t]
\centerline{
\includegraphics[height=5cm]{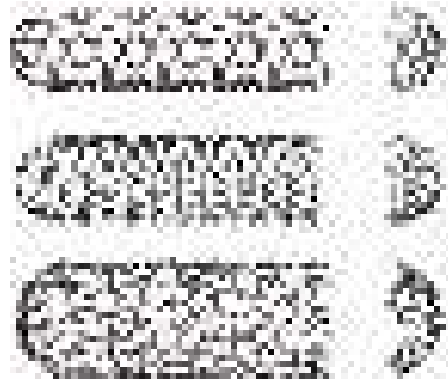} \hspace{1cm}}
\caption{The fullerene C$_{60}$ and spheroidal ones as a cap at the
end of some nanotubes.} \label{fig2}
\end{figure}

The high flexibility of carbon allows producing variously shaped
carbon nanostructures: fullerenes, nanotubes, nanohorns, cones,
toroids, graphitic onions, etc. Some of them as e.g.nanotubes are
composition of tubes and fullerenes (as in
Fig.~\ref{fig1}).Historically, the fullerenes C$_{60}$ (nicknamed
also as Buckminsterfullerene or 'bucky ball') were first discovered
in 1985~\cite{kroto}. They are tiny molecular cages of carbon having
60 atoms and making up the mathematical shape called a truncated
icosahedron (12 pentagons and 20 hexagons). The amount of C$_{60}$
had been produced in small quantities  until 1990 when an efficient
method of production was achieved~\cite{kratschmer}. Since then,
there were produced variously shaped fullerene molecules. These
molecules are formed when $12$ pentagons are introduced in the
hexagonal lattice of graphene. Soon after the fullerenes, the carbon
nanotubes of different diameters and helicity~\cite{iijima} were
produced. We can also image the construction of nanotube with
fullerenes as a cap at the end of nanotubes (see Fig.~\ref{fig2}).
The mechanical, magnetic, and especially electronic properties of
carbon nanotubes are found to be very specific (see,
e.g.,~\cite{ebbesen}).

\begin{figure}[t]
\centerline{
\includegraphics[height=7cm]{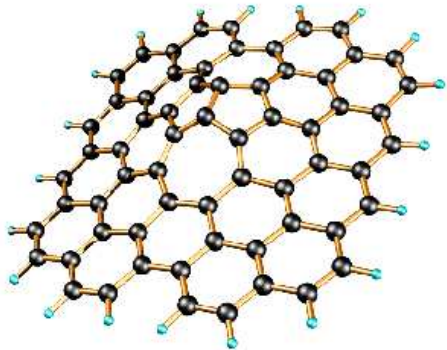} \hspace{1cm}}
\caption{The pentagonal and heptagonal defects in the hexagonal
network.} \label{fig3}
\end{figure}

The fullerenes are composed of hexagonal and pentagonal carbon
rings. However, the structures having heptagonal rings are also
possible. A junction which connect carbon nanotubes with different
diameters through a region sandwiched by pentagon-heptagon pair has
been observed in the transmission electron
microscope~\cite{Iijima1}. A pair of five- and seven-membered rings
have to be imposed to connect two different types of carbon
nanotubes~\cite{Dress}. Theoretically the closed fullerenes and
nanotubes exhibiting high topologies (from genus 5 to genus 21) were
suggested in~\cite{terrones}. This follows from the known Euler's
theorem that relates the number of vertices, edges and faces of an
object. For the hexagonal carbon lattice it can be written in the
form~\cite{terrones}
\begin{equation}
...2n_4+n_5-n_7-2n_8... = \sum(6-x)n_x = \chi= 12(1-g),
\label{eq2.1}
\end{equation}
where $n_x$ is the number of polygons having $x$ sides, $\chi$ is
the Euler characteristic which is a geometrical invariant related
to the topology of the structure, and  $g$ is the genus or a
number of handles of an arrangement. According to (\ref{eq2.1})
there is no contribution to the Gaussian curvature for $x=6$. This
means that two-dimensional carbon lattice consisting only of
hexagons is flat. By their nature, pentagons (as well as other
polygons with $x\neq 6$) in a graphite sheet are topological
defects (see Fig.~\ref{fig3}).
In particular, fivefold coordinated particles are orientational
60$^{\circ}$ disclination defects in the sixfold coordinated
triangular lattice. Description of the electronic structure requires
formulating a theoretical model describing electrons on arbitrary
curved surfaces with disclinations taken into account. The most
important fact found in~\cite{mele,Wallace,Weiss} is that the
electronic spectrum of a single graphite plane linearized around the
corners of the hexagonal Brillouin zone coincides with that of the
Dirac equation in (2+1) dimensions. This finding stimulated a
formulation of some field-theory models for Dirac fermions on
hexatic surfaces to describe electronic structure of variously
shaped carbon materials: fullerenes~\cite{jetplet00,jose92} and
nanotubes~\cite{kane}. The field-theory models for Dirac fermions on
a plane and on a sphere~\cite{jose91} were invoked to describe
variously shaped carbon materials.

\chapter[Electronic Structure of
Spheroidal Fullerenes]{Electronic properties} \label{chap:Electro}

Carbon nanoparticles and their electronic properties have been a
subject of an extensive study. Electronic states in nanotubes,
fullerenes, nanocones, nanohorns as well as in other carbon
configurations are the subject of an increasing number of
experimental and theoretical studies. It has been predicted and
later observed in experiment that bending or stretching a nanotube
change its band structure changing therefore the electrical
properties: stretched nanotubes become either more or less
conductive. Carbon  nanotubes can be either a metal or
semiconductor, depending on their diameters and helical
arrangement~\cite{ando1}. This finding could allow to build
nanotube-based transducers sensitive to tiny forces.

The electronic properties depends on the topological defects. The
peculiar electronic states due to topological defects have been
observed in different kinds of carbon nanoparticles by scanning
tunneling microscopy (STM). For example, STM images with five-fold
symmetry (due to pentagons in the hexagonal graphitic network)
have been obtained in the C$_{60}$ fullerene molecule~\cite{hou}.
The peculiar electronic properties at the ends of carbon nanotubes
(which include several pentagons) have been probed experimentally
in~\cite{carroll,kim}. Recently, the electronic structure of a
single disclination has been revealed on an atomic scale by
STM~\cite{an}.

The problem of peculiar electronic states near the pentagons in
curved graphite nanoparticles was the subject of intensive
theoretical studies in fullerenes~\cite{jose92,jetplet00},
nanotubes~\cite{kane}, nanohorns~\cite{berber}, and
cones~\cite{jetp03}. In particular, an analysis within the
effective-mass theory has shown that a specific
$\sqrt{3}\times\sqrt{3}$ superstructure induced by pentagon
defects can appear in nanocones~\cite{kobayashi}. This prediction
has been experimentally verified in~\cite{an}. A recent
study~\cite{charlier} within both tight-binding and {\it ab
initio} calculations shows a presence of sharp resonant states in
the region close to the Fermi energy. The localized cap states in
nanotubes have been recently studied in~\cite{yaguchi}.

It is interesting to note that the problem of specific electronic
states at the Fermi level due to disclinations is similar to that of
the fermion zero modes for planar systems in a magnetic field.
Generally, zero modes for fermions in topologically nontrivial
manifolds have been of current interest both in the field theory and
condensed matter physics. As was revealed, they play a major role in
getting some insight into understanding anomalies~\cite{jackiw77}
and charge fractionalization that results in unconventional
charge-spin relations (e.g. the paramagnetism of charged
fermions)~\cite{jackiw81} with some important implications for
physics of superfluid helium (see, e.g., review~\cite{salomaa}).
$3D$ space-time Dirac equation for massless fermions in the presence
of the magnetic field was found to yield $N-1$ zero modes in the
N-vortex background field~\cite{jackiw84}. As it was shown
in~\cite{jetplet00}, the problem of the local electronic structure
of fullerene is closely related to Jackiw's
analysis~\cite{jackiw84}. An importance of the fermion zero modes
was also discussed in the context of the high-temperature chiral
superconductors~\cite{volovik99b}. Among different theoretical
approaches to describe graphene compositions, the continuum model
play a special role. It allow us to describe some integral features
of similar carbon structures. It should be noted that the
formulation of any continuum model begins from description of a
graphite sheet (graphene). We briefly report now the effective mass
theory for graphene~\cite{Wallace,Weiss}. A single layer of graphite
forms a two-dimensional material. Two Bloch functions constructed
from atomic orbitals for two inequivalent carbon atoms, A and B, in
the unit cell provide the basis function for graphene. The Bloch
basis functions $\Phi _{j}$ ($j=A,B$) are constructed from atomic
$2p_{z}$ orbitals
\begin{equation}
\Phi _{A}(\vec{k},  \vec{r})=\sum _{n} \frac{1}{\sqrt{N}}\ e^{i
\vec{k}\  \vec{r}_{n}} f( \vec{r}- \vec{r_{n}}),
\end{equation}
\begin{equation}
\Phi _{B}(\vec{k},  \vec{r})=\sum _{n} \frac{1}{\sqrt{N}}\ e^{i
\vec{k}\  \vec{r}_{n}} f( \vec{r}- \vec{d}-\vec{r_{n}}),
\end{equation}
$\vec{d}$ is a coordinate of $B$ atom in the unit cell, $N$ is a
number of unit cell, $ \vec{r}_{n}$ is a unit cell coordinate, $f$
is $2p_{z}$ orbital, $ \vec{k}$ is the wave vector. Now we want to
solve the Schr\"{o}dinger equation
\begin{equation}
\hat{H}\Phi =E \Phi,
\end{equation}
where $ \Phi =c_{A} \Phi_{A}+c_{B} \Phi_{B}$ and $ \hat{H}$ is the
Hamiltonian of the system. We get the secular equations
$$H_{AA} c_{A}+H_{AB} C_{B}=E C_{A},$$
\begin{equation}
H_{BA} c_{A}+H_{BB} C_{B}=E C_{B},
\end{equation}
here
\begin{equation}
H_{ij}=\langle \Phi_{i}| \hat{H}|\Phi_{j} \rangle, \quad i,j=A,B.
\end{equation}
In a tight-binding approximation it have a form
$$\hat{H}_{AA}=\hat{H}_{BB}=\varepsilon _{2p},$$
\begin{equation}
\hat{H}_{AB}=t\left( e^{i \vec{k} \vec{\tau}_{1}} + e^{i \vec{k}
\vec{\tau}_{2}}+e^{i \vec{k} \vec{\tau}_{3}}\right),
\end{equation}
where $\varepsilon_{2p}$ is a orbital energy of the $2p_{z}$ level
and $\vec{\tau}_{i}$ are the coordinates of the three
nearest-neighbor $B$ atoms relative to an A atom
\begin{equation}
t= \langle f( \vec{r})| \hat{H}|f( \vec{r}- \vec{d}\ )\rangle.
\end{equation}
Solving the secular equation
\begin{equation}
det\left( H-E\right)=0,
\end{equation}
the eigenvalues $E( \hat{k})$ are obtained in the form
\begin{equation}
E( \vec{k})=\varepsilon _{2p} \pm t
\sqrt{1+4\cos\frac{ak_{x}}{2}\cos \frac{\sqrt{3}ak_{y}}{2}+4\cos^{2}
\frac{ak_{x}}{2}}.
\end{equation}
We have
\begin{equation}
E( \vec{K})=E( \vec{K}^{'})=\varepsilon _{2p},
\end{equation}
and the $ \vec{K}$ and $ \vec{K}^{'} $ points are the corners of the
Brillouin zone. In $ \vec{k}. \vec{p}$ theory we approximate the
wave function at wave vector $\hat{k}=\hat{K}+\hat{\kappa}$ by
\begin{equation}
\Phi(\hat{k},\hat{r}) =c_{A}(\hat{\kappa})e^{i\hat{\kappa}.\hat{r}}
\Phi_{A}(\hat{K},\hat{r})+c_{B}(\hat{\kappa})
e^{i\hat{\kappa}.\hat{r}}\Phi_{B}(\hat{K},\hat{r}).
\end{equation}
Inserting $\Phi$ into Schr\"{o}dinger equation, keeping terms of
order $\vec{\kappa}$, we get the secular equation ($\hbar=c=1$
units are used)
\begin{equation}
\frac{\vec{\kappa}}{m}.\left(\begin{array}{cc}\vec{p}_{AA}&\vec{p}_{AB}\\
\vec{p}_{BA}&\vec{p}_{BB}\end{array}\right)\left(\begin{array}{c}c_{A}(\vec{\kappa})\\
c_{B}(\vec{\kappa})\end{array}\right)=E(\vec{\kappa})\left(\begin{array}{c}c_{A}(\vec{\kappa})\\
c_{B}(\vec{\kappa})\end{array}\right),
\end{equation}
where $E(\vec{\kappa})=E(\vec{k})-E(\vec{K})$ and
\begin{equation}
\vec{p}_{ij}=\int
\Phi_{i}^{*}(\hat{K},\hat{r})\vec{p}\Phi_{j}(\hat{K},\hat{r})d\vec{r}
\ i,j=A,B.
\end{equation}
It can be shown from group theoretical argument (and can be verify
directly within the one-orbital tight-binding model) that the
secular equation can be write in the form~\cite{Weiss,mele}
\begin{equation}
v_{F}\left(\begin{array}{cc}0&\kappa_{x}-i\kappa_{y}\\
\kappa_{x}+i\kappa_{y}&0\end{array}\right)\left(\begin{array}{c}c_{A}(\vec{\kappa})\\
c_{B}(\vec{\kappa})\end{array}\right)=E(\vec{\kappa})\left(\begin{array}{c}c_{A}(\vec{\kappa})\\
c_{B}(\vec{\kappa})\end{array}\right),
\end{equation}
where $v_{F}=\frac{\sqrt{3}at}{2}$, $a$ is the lattice constant
and
\begin{equation}
E(\vec{\kappa})=\pm v_{F}|\vec{\kappa}|.
\end{equation}
The upper half of the energy dispersion curves describes the
$\pi^{*}$-energy anti-bonding band, and the lower half is the $\pi$
energy bonding band. Since there are two $\pi$ electrons per unit
cell, the lower $\pi$ band is fully occupied. The points $K$,$K^{'}$
create the Fermi surface of the graphene. When an external gauge
fields are imposed, the translation symmetry can be broken and
eigenfunction can no longer be labeled by $\vec{\kappa}$. This
requires a generalization of the trial wave function
\begin{equation}
\Phi(\hat{r}) =\int c_{A}(\hat{\kappa})e^{i\hat{\kappa}.\hat{r}}
\Phi_{A}(\hat{K},\hat{r}) d\vec{\kappa}+\int c_{B}(\hat{\kappa})
e^{i\hat{\kappa}.\hat{r}}\Phi_{B}(\hat{K},\hat{r}) d\vec{\kappa}.
\end{equation}
Inserting this function into Schr\"{o}dinger equation leads to an
equation~\cite{mele,lutt}
\begin{equation}
\left(-iv_{F}\vec{\sigma}.(\vec{\nabla}-i\vec{W})-E\right)
\left(\begin{array}{c}\Psi_{A}(\vec{r})\\
\Psi_{B}(\vec{r})\end{array}\right) =0,
\end{equation}
here $\vec{\sigma}=(\sigma_{x},\sigma_{y})$ are conventional Pauli
matrices, $\vec{W}$ is the U(1) external gauge field and
$\Psi_{A}(\vec{r})$, $\Psi_{B}(\vec{r})$ are envelope functions,
smoothly varying in the scale of the lattice constant $a$,
multiplying graphene Bloch function
\begin{equation}
\Phi(\hat{r})
=\Psi_{A}(\hat{r})\Phi_{A}(\hat{K},\hat{r})+\Psi_{B}(\hat{r})
\Phi_{B}(\hat{K},\hat{r}),
\end{equation}
\begin{equation}
\Psi_{i}(\hat{r})=\int
c_{i}(\hat{\kappa})e^{i\hat{\kappa}.\hat{r}} d\vec{\kappa} \ \ \
 \ \ i=A,B.
\end{equation}
Eq.(2.17) is algebraically identical to a two-dimensional Dirac
equation, where components of spinor represent localization of
electron on the graphene sublattice A and sublattice B,
respectively. We can get the similar equation for envelope functions
near $K^{'}$ point. Formally Eq.(2.17) can be obtained from
Eq.(2.14) by exchanging $\kappa_{k}\rightarrow
-i(\partial_{k}-iW_{k})$, $c_{A(B)}\rightarrow \Psi_{A(B)}(\vec{r})$
and $E(\vec{\kappa})\rightarrow E$.

\chapter[Electronic Structure of
Spheroidal Fullerenes]{Theory} \label{chap:Theory}

Now we will formulate a theoretical model describing electrons on
arbitrary curved surfaces with disclinations taken into account to
investigate the electronic structure of fullerenes. As it was
mention in the previous section the electronic spectrum of a single
graphite plane linearized around the corners of the hexagonal
Brillouin zone coincides with that of the Dirac equation in (2+1)
dimensions. This finding stimulated formulation of some field-theory
models for Dirac fermions on hexatic surfaces to describe electronic
structure of variously shaped carbon materials:
fullerenes~\cite{jetplet00}, nanotubes~\cite{kane,yaguchi}, and
cones~\cite{lammert,jetplet01}.

The effective-mass theory for a graphene sheets can be generalized
to graphene surfaces containing five carbon atom rings and creating
geometrical structures as a cones or fullerenes. In the continuum
description pentagonal ring are described as a localized fictitious
gauge fluxes. In such structures there is no globally coherent
distinction between $K$ and $K^{'}$ points.

Following the ideas above and describe fermions in a curved
background, we need a set of orthonormal frames $\{e_{\alpha}\}$
which yield the same metric, $g_{\mu\nu}$, related to each other by
the local $SO(2)$ rotation,
$$e_{\alpha}\to e'_{\alpha}={\Lambda}_{\alpha}^{\beta}e_{\beta},\quad
{\Lambda}_{\alpha}^{\beta}\in SO(2).$$ It then follows that
$g_{\mu\nu} = e^{\alpha}_{\mu}e^{\beta}_{\nu} \delta_{\alpha \beta}$
where $e_{\alpha}^{\mu}$ is the zweibein, with the orthonormal frame
indices being $\alpha,\beta=\{1,2\}$, and coordinate indices
$\mu,\nu=\{1,2\}$. As usual, to ensure that physical observables are
independent of a particular choice of the zweinbein fields, a local
$so(2)$ valued gauge field $\omega_{\mu}$ must be introduced. The
gauge field of the local Lorentz group is known as the spin
connection. For the theory to be self-consistent, the zweibein
fields must be chosen to be covariantly constant~\cite{Nakahara}:
$${\cal D}_{\mu}e^{\alpha}_{\nu}:=\partial_{\mu}e^{\alpha}_{\nu}
-\Gamma^{\lambda}_{\mu\nu}e^{\alpha}_{\lambda}+(\omega_{\mu})^{\alpha}_{\beta}
e^{\beta}_{\nu}=0,$$ which determines the spin connection
coefficients explicitly
\begin{equation}
(\omega_{\mu})^{\alpha\beta}= e_{\nu}^{\alpha}D_{\mu}e^{\beta\nu}.
\label{eq3.6}
\end{equation}
Finally, the Dirac equation on a surface $\Sigma$ in the presence of
the $U(1)$ external gauge field $W_{\mu}$ is written
as~\cite{Davies}

\begin{equation}
i\gamma^{\alpha}e^{\mu}_{\alpha}\left(\nabla_{\mu}-ia_{\mu}-iW_{\mu}\right)\psi(\vec
r)=E\psi(\vec r).\label{eq3.1}
\end{equation}
where $\nabla_{\mu}=\partial_{\mu}+\Omega_{\mu}$ with
\begin{equation}
\Omega_{\mu}=\frac{1}{8}\omega^{\alpha\ \beta}_{\ \mu}
[\gamma_{\alpha},\gamma_{\beta}], \label{eq3.8}
\end{equation}
being the spin connection term in the spinor representation and
\begin{equation}
\gamma^\alpha=-I
\sigma_{\alpha}=-\left(\begin{array}{cc}\sigma_{\alpha}&0\\0&\sigma_{\alpha}\end{array}\right)
\end{equation}
\begin{equation}
\oint a_{\mu}dx^{\mu}=\frac{2 \pi}{4}\tau_{2}I.
\end{equation}
The matrix $\tau_{2}$ acting in $K$-spin space. The energy $E$ is
measured relative to the Fermi energy, and the spinor
$\psi=(\Psi^{K}_{A}\Psi^{K}_{B}\Psi^{K^{'}}_{A}\Psi^{K^{'}}_{B})^{T}$
where $\Psi( \vec{r})$ are envelope functions. For our purpose, we
need incorporating both a disclination field and a nontrivial
background geometry. A possible description of disclinations on
arbitrary two-dimensional elastic surfaces is offered by the gauge
approach~\cite{jpa99}. In accordance with the basic assumption of
this approach, disclinations can be incorporated in the elasticity
theory Lagrangian by introducing a compensating $U(1)$ gauge fields
$W_{\mu}$. It is important that the gauge model admits exact
vortex-like solutions for wedge disclinations~\cite{jpa99} thus
representing a disclination as a vortex of elastic medium. The
physical meaning of the gauge field is that the elastic flux due to
rotational defect (that is directly connected with the Frank vector
(see the next section)) is completely determined by the circulation
of the $W_{\mu}$ field around the disclination line. In the gauge
theory context, the disclination field can be straightforwardly
incorporated in~(\ref{eq3.1}) by the standard substitution
$\partial_{\mu}=\partial_{\mu}-iW_{\mu}$.

Within the linear approximation to gauge theory of disclinations
(which amounts to the conventional elasticity theory with linear
defects), the basic field equation that describes the $U(1)$ gauge
field in a curved background is given by
\begin{equation} D_{\mu}F^{\mu k}=0,\quad
F^{\mu k}=\partial^{\mu}W^k-\partial^kW^{\mu}, \label{eq3.2}
\end{equation}
where covariant derivative $D_{\mu}:=\partial_{\mu}+\Gamma_{\mu}$
involves the Levi-Civita (torsion-free, metric compatible)
connection
\begin{equation}
\Gamma^k_{\mu\lambda}:=(\Gamma_{\mu})^k_{\lambda} =
\frac{1}{2}g^{kl} \left(\frac{\partial g_{l\lambda}}{\partial
x^{\mu}}+\frac{\partial g_{\mu l}} {\partial
x^{\lambda}}-\frac{\partial g_{\mu\lambda}}{\partial x^l}\right),
\label{eq3.3}
\end{equation}
with $g_{\mu k}$ being the metric tensor on a Riemannian surface
$\Sigma$ with local coordinates $x^{\mu}=(x^1,x^2)$. For a single
disclination on an arbitrary elastic surface, a singular solution
to~(\ref{eq3.2}) is found to be~\cite{jpa99}
\begin{equation}
W^k = -\nu\varepsilon^{k\lambda}D_{\lambda} G(x,y), \label{eq3.4}
\end{equation}
where
\begin{equation}
D_{\mu}D^{\mu}G(x^1,x^2) = 2\pi \delta^2(x^1,x^2)/\sqrt g,
\label{eq3.5}
\end{equation}
with $\varepsilon_{\mu k}=\sqrt{g}\epsilon_{\mu k}$ being the fully
antisymmetric tensor on $\Sigma,\quad
\epsilon_{12}=-\epsilon_{21}=1.$ It should be mentioned that
eqs.~(\ref{eq3.2}-\ref{eq3.5}) self-consistently describe a defect
located on an arbitrary surface~\cite{jpa99}.

The general analytical solution to~(\ref{eq3.1}) is known only for
chosen geometries. One of them is the cone~\cite{lammert,jetplet01}.
For the sphere which are of interest here, there were used some
approximations. In particular, asymptotic solutions at small $r$
(which allow us to study electronic states near the disclination
line) were considered in~\cite{jetp03}. For this reason, also the
numerical calculations were performed in~\cite{pincak}. The results
of both analytical and numerical studies of spherical geometry will
be presented in the next section.

\chapter[Electronic Structure of
Spheroidal Fullerenes]{Spherical molecules} \label{chap:Spheric}

According to the Euler's theorem, the fullerene molecule consists of
twelve disclinations. Generally, it is difficult to take into
account properly all the disclinations. There are two ways to
simplify the problem. The first one is considering a situation near
a single defect. This approximation we briefly sketch in this
chapter. The second possibility is replacing the fields of twelve
disclinations by the fields of the magnetic monopole. This
approximation will be described in the next section.

\section[]{The Model}
We start the section with the computation of disclination fields. It
will be used as a gauge fields in Dirac equation on the sphere. To
describe a sphere, we employ the polar projective coordinates
$x^1=r,x^2=\varphi;\, 0\le r<\infty, 0\le\varphi<2\pi$ with $R$
being the radius of the sphere. In these coordinates, the metric
tensor becomes
\begin{equation}
g_{rr}=4R^4/(R^2+r^2)^2,\, g_{\varphi\varphi}=4R^4r^2/(R^2+r^2)^2,\,
g_{r\varphi}= g_{\varphi r}=0, \label{eq4.1}
\end{equation}
so that $$\sqrt g:=\sqrt{\det{||g_{\mu\nu}||}}=4R^4r/(R^2+r^2)^2.$$
Nonvanishing connection coefficients (\ref{eq3.3}) take the form
$$\Gamma^r_{rr}=-\frac{2r}{R^2+r^2},\quad \Gamma^r_{\varphi\varphi}=
-r\frac{R^2-r^2}{R^2+r^2},\quad \Gamma^{\varphi}_{r\varphi}=
\frac{1}{r}\frac{R^2-r^2}{R^2+r^2},$$ and the general representation
for the zweibeins is found to be
$$e^1_{\ r} = e^2_{\ \varphi}=2R^2\cos\varphi/(R^2+r^2),\, e^1_{\ \varphi} =
-e^2_{\ r}=-2R^2\sin\varphi/(R^2+r^2),$$ which in view
of~(\ref{eq3.6}) gives
\begin{equation}
\omega^{12}_r=\omega^{21}_r= 0,\quad \omega^{12}_{\varphi}=
-\omega^{21}_{\varphi}= 2r^2/(R^2+r^2)=:2\omega.
\label{eq4.2}\end{equation}

From the solutions (\ref{eq3.4}) and (\ref{eq3.5}) one can easily
found
$$ G=\log r;\quad W_r=0,\quad W_{\varphi}=\nu,\quad r\neq 0.$$
It describes locally a topological vortex on the Euclidean plane,
which confirms the observation that disclinations can be viewed as
vortices in elastic media. The elastic flux is characterized by the
Frank vector $\vec\omega$, $|\vec\omega|=2\pi\nu$ with $\nu$ being
the Frank index and the elastic flow through a surface on the sphere
is given by the circular integral
$$\frac{1}{2\pi}\oint\vec{W}d\vec{r} =
\nu.$$  There are no restrictions on the value of the winding number
$\nu$ apart from $\nu>-1$ for topological reasons. This means that
the elastic flux is 'classical' in its origin; i.e., there is no
quantization (in contrast to the magnetic vortex). If we take into
account the symmetry group of the underlying crystal lattice, the
possible values of $\nu$ become 'quantized' in accordance with the
group structure (e.g., $\nu =1/6, 1/3, 1/2, ...$ for the hexagonal
lattice). We define $\nu=n/6$, where $n$ is the number of pentagons
at the apices. It is interesting to note that in some physically
interesting applications vortices with the fractional winding number
have already been considered.  A detailed theory of magnetic
vortices on a sphere has been presented in~\cite{ovrut}.

The Dirac matrices in $2D$ space,can be chosen as the Pauli
matrices: $\gamma^1=-\sigma^2, \gamma^2=\sigma^1$ and (\ref{eq3.8})
reduces to
\begin{equation}
\Omega_{\varphi}=i\omega\sigma^3. \label{eq4.3}
\end{equation}
From the assumptions above the Dirac operator $\hat{\cal
D}:=i\gamma^{\alpha}e_{\alpha}^{\ \mu}(\nabla_{\mu}+iW_{\mu})$ on
the two-sphere becomes
\begin{eqnarray}
\hat{\cal D}={\hat{\cal D}}^{\dagger}=
\frac{r^2+R^2}{2R^2}\left[\matrix{0 & e^{-i\varphi}(-\partial_r+
\frac{i\partial_{\varphi}+\nu}{r}+\frac{\omega}{r})\cr
e^{i\varphi}(\partial_r+\frac{i\partial_{\varphi}+\nu}{r}-\frac{\omega}{r})&0}
\right]. \label{eq4.4}\end{eqnarray}

The generator of the local Lorentz transformations
${\Lambda}_{\alpha}^{\beta}\in SO(2)$ takes the form
$-i\partial_{\varphi}$, whereas the generator of the Dirac spinor
transformations $\rho(\Lambda)$ is
$$\Sigma_{12}=\frac{i}{4}[\gamma_1,\gamma_2]=\frac{1}{2}\sigma^3.$$
For massless fermions $\sigma^3$ serves as a conjugation matrix, and
the energy eigenmodes are symmetric with respect to $E=0$
($\sigma^3\psi_E=\psi_{-E}$). The total angular momentum of the $2D$
Dirac system is therefore given by
$$L_z=-i\partial_{\varphi} +\frac{1}{2}\sigma^3,$$
which commutes with the operator~(\ref{eq4.4}). Accordingly, the
eigenfunctions are classified with respect to the eigenvalues of
$J_z=j+1/2,\,\, j=0,\pm 1,\pm 2,...,$ and are to be taken in the
form
\begin{equation}
\psi = \left(u(r)e^{i\varphi j}\atop{v(r)e^{i\varphi (j+1)}}\right).
\label{eq4.5}\end{equation}

From the (\ref{eq4.4}) is follows that the spin connection term can
be taken into account by redefining the wave function as
\begin{equation}
\psi=\tilde{\psi}\sqrt{R^2+r^2}, \label{eq4.6}
\end{equation}
which reduces eigenvalue problem~(\ref{eq3.1}) to
\begin{eqnarray}
\partial_{r}\tilde u-\frac{(j-\nu)}{r}\tilde u=\tilde{E}\tilde v,\nonumber \\
-\partial_{r}\tilde v-\frac{(j+1-\nu)}{r}\tilde v=\tilde{E}\tilde u,
\label{eq4.7}
\end{eqnarray}
where $\tilde{E}=2R^2 E/(R^2+r^2)$.

\section[]{Extended Electron States}

We will consider an approximate solution to (\ref{eq4.7}). Because
we are interested in electronic states near the disclination line,
we can restrict our consideration to the case of small $r$. In this
case, a solution to (\ref{eq4.7}) (with (\ref{eq4.6}) taken into
account) is found to be
\begin{equation}
\left(u\atop{v}\right) = A\left(J_{\eta}(2Er)\atop{\pm
J_{\bar\eta}(2Er)}\right) \label{eq4.8}
\end{equation}
where $\eta=\pm(j-\nu)$, $\bar\eta=\pm(j-\nu+1)$, and $A$ is a
normalization factor. Therefore, there are two independent solutions
with $\eta(\bar\eta)>0$ and $\eta(\bar\eta)<0$. Notice that
respective signs $\pm$ in (\ref{eq4.8}) correspond to states with
$E>0$ and $E<0$. As already noted, $\sigma^3$ serves as the
conjugation matrix for massless fermions, and the energy eigenmodes
are symmetric with respect to $E=0$. One can therefore consider
either case, for instance, $E>0$.

The important restrictions come from the normalization condition
\begin{equation}
\int(|u|^2+|v|^2)\sqrt{g}dx^1dx^2 = 1. \label{eq4.9}
\end{equation}
From (\ref{eq4.8}), it follows that $A^2\sim E$. On the other hand,
the integrand in (\ref{eq4.9}) must be nonsingular at small $Er$.
This imposes a restriction on possible values of $j$. Namely, for
$\eta,\bar\eta>0$ one obtains $j-\nu>-1/2$, and for
$\eta,\bar\eta<0$ one has $j-\nu<-1/2$. As is seen, possible values
of $j$ do not overlap at any $\nu$.

In the vicinity of a pentagon, the electron wave function reads
\begin{equation}
\left(u\atop v\right) \sim \left(E^{1/2+\eta}r^{\eta} \atop
E^{1/2+\bar\eta}r^{\bar\eta}\right). \label{eq4.10}
\end{equation}
In particular, in the leading order, one obtains $\Psi\sim \sqrt{E},
\ \Psi\sim E^{1/3}r^{-1/6}$, and $\Psi\sim E^{1/6}r^{-1/3}$ for
$\nu=0,\ 1/6,\  1/3$, respectively. Because the local density of
states diverges as $r\rightarrow 0$, it is more appropriate to
consider the total density of states on a patch $0<r\leq\delta$ for
small $\delta$, rather than the local quantities. For this, the
electron density should be integrated over a small disk $|r|<\delta$
(recall that $r,\varphi$ are stereographically projected coordinates
on the sphere). The result is
\begin{eqnarray}
D(E,\delta) \propto \left\{\begin{array}{ll}
(E\delta)\delta , &\nu=0; \\[2mm]
(E\delta)^{2/3}\delta , &\nu=1/6, 5/6; \\[2mm]
(E\delta)^{1/3}\delta,  &\nu=1/3, 2/3;\\[2mm]
\delta,        &\nu=1/2;\\[2mm]
\end{array}\right.
\label{eq4.11}
\end{eqnarray}
For the defect free case ($\nu=0$) we obtain the well-known behavior
of the total density of electronic states (DOS) in the $\delta$ disk
given by $D(E,\delta)\sim E\delta^2$ (in accordance with the
previous analysis~\cite{mele}). For $\nu=1/6, 1/3, 2/3, 5/6$, the
low-energy total DOS has a cusp which drops to zero at the Fermi
energy. Most intriguing is the case where $\nu=1/2$ and a region of
a nonzero DOS across the Fermi level is formed. This implies local
metallization of graphite in the presence of $180^\circ$
disclination. In the fullerene molecule, however, there are twelve
$60^\circ$ disclinations, and therefore, the case $\nu=1/6$ is
actually realized.

\section[]{Numerical Results}

\begin{figure}
\centering
\includegraphics[scale=0.8] {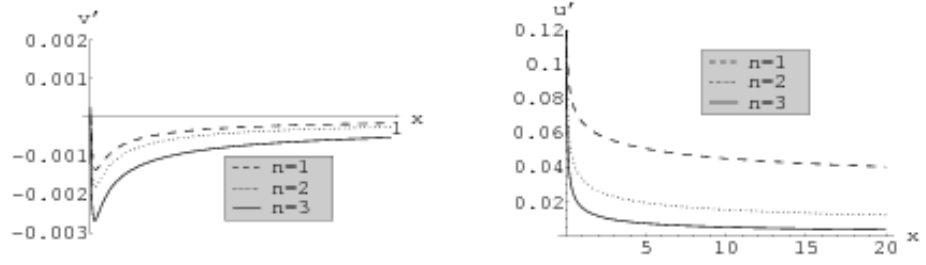}
\caption{The solutions $v'(x) ,u'(x) $  for different $n$.}
\label{fig4}
\end{figure}

\begin{figure*}[t]
\centering
\includegraphics[scale=.4] {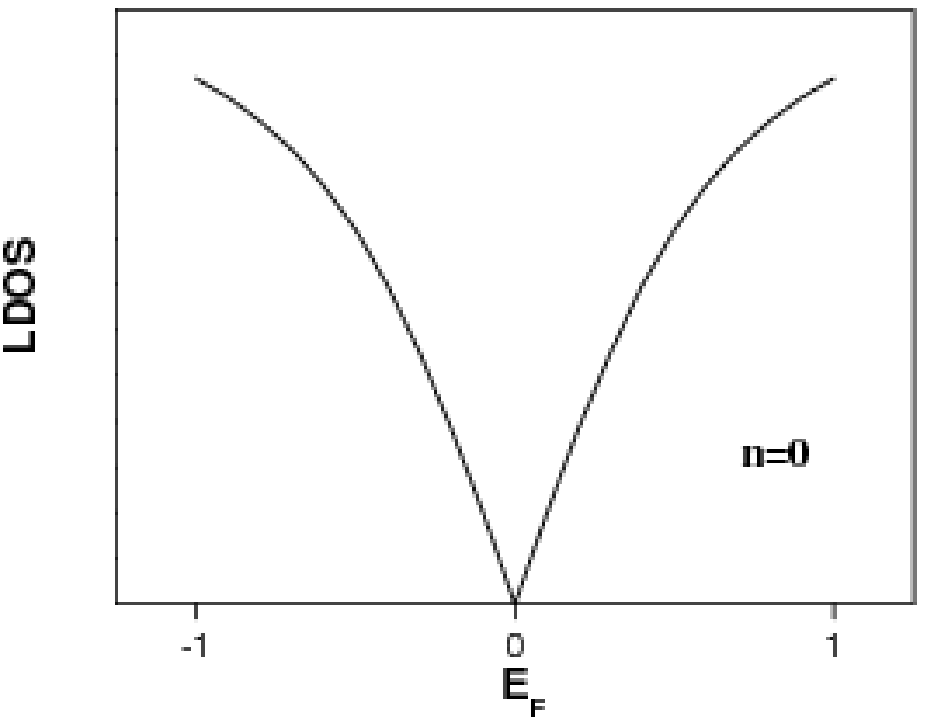}
\includegraphics[scale=.4] {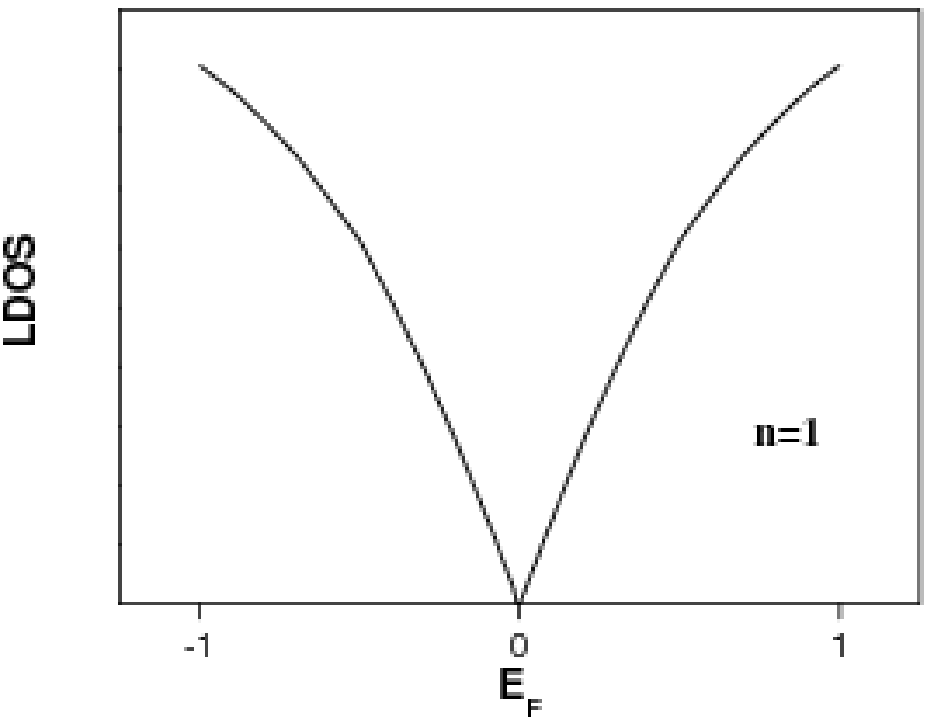}\\
\includegraphics[scale=.4] {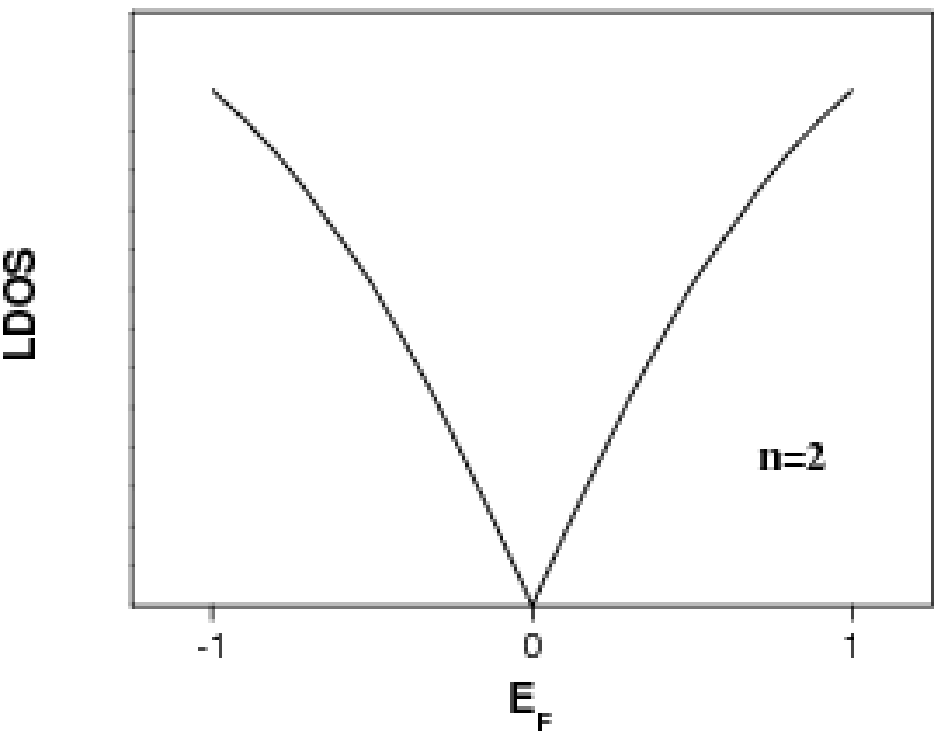}
\includegraphics[scale=.4] {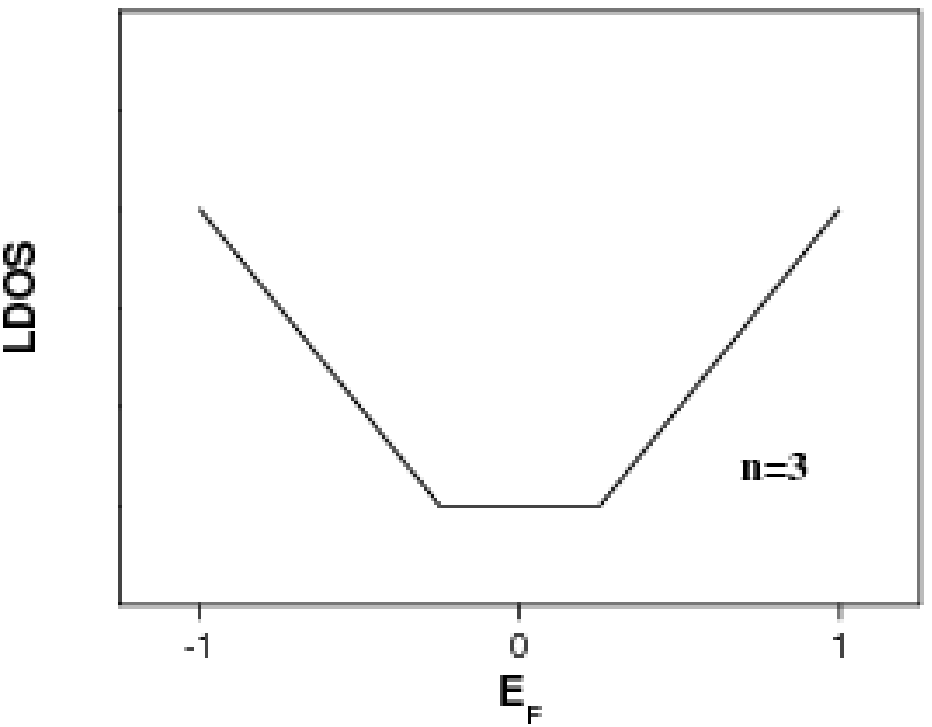}
\caption{Schematic densities of states near the Fermi energy in the
case of sphere.}
\label{fig5}
\end{figure*}

\begin{figure*}[t]
\centering
\includegraphics[height=3.5cm] {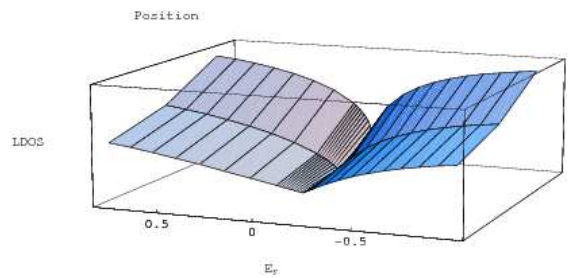}
\caption{3d schematic plotting of the DOS near the Fermi energy for
$n=0,1,2$ (going from the front side to the back side).}
\label{fig6}
\end{figure*}

The numerical calculations for the different type of carbon
nanostructures as well as for the case of sphere were presented in
paper~\cite{pincak}. As a starting point, the analytical asymptotic
solutions found in the previous section are considered. The initial
value of the parameter $r$ is defined as $r=10^{-4}$. It is worth
noting that the choice of the boundary conditions does not influence
the behavior of the calculated wave functions and only the starting
point depends on it. A dimensionless substitution $x=Er$ is used.
The normalized numerical solutions to (\ref{eq4.7}) are given in
Fig.~\ref{fig4}. The parameters are chosen to be $E=0.01$ and $R=1$.
Notice that here we present the solutions for dotted values
$v'(=\tilde v)$ and $u' (=\tilde u)$. The local DOS is shown
schematically in Figs.~\ref{fig5},~\ref{fig6} for different $n$. The
Fig.~\ref{fig6} describes also the dependence of the local DOS on a
position of the maximum value of integrand in (\ref{eq4.9})(which
actually characterizes the numerically calculated localization point
of an electron).
Here and below $\delta=0.1$. Notice that in fact the choice of the
value of $\delta$ does not influence the characteristic behavior of
LDOS. As is seen, the DOS has a cusp which drops to zero at the
Fermi energy. The case $n=3$ becomes distinguished. Let us emphasize
once more that in the fullerene molecule there are twelve $60^\circ$
disclinations, so that the case $n=1$ is actually realized.
\newpage

\chapter[Electronic Structure of
Spheroidal Fullerenes]{Spheroidal geometry approach to fullerenes}
\label{chap:Spheroidal}

\begin{figure*}[t]
\centering
\includegraphics[height=7cm] {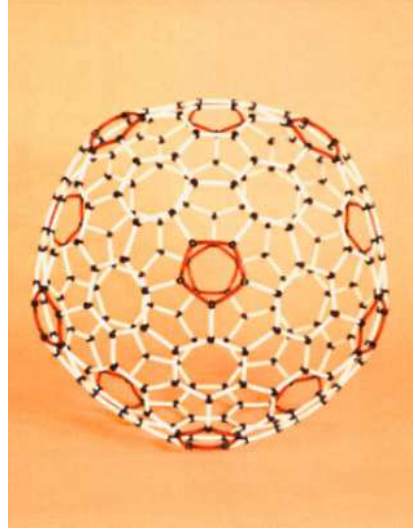}
\caption{Picture of molecular model of giant spheroidal fullerene
$C_{240}$.} \label{fig7}
\end{figure*}

\begin{figure*}[t]
\centering
\includegraphics[height=7cm] {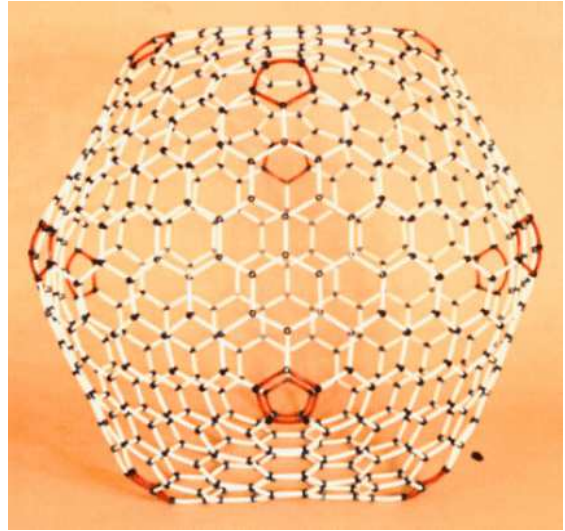}
\caption{Picture of molecular model of giant spheroidal fullerene
$C_{540}$.} \label{fig8}
\end{figure*}

Geometry, topological defects and the peculiarity of graphene
lattice have a pronounced effect on the electronic structure of
fullerene molecules. The most extensively studied $C_{60}$ molecule
is an example of a spherical fullerene nicknamed a "soccer
ball"~\cite{kroto1}. Moreover, the spherical fullerenes as $C_{60}$
are stable towards fragmentation than the other bigger
fullerenes~\cite{Ryan}. The family of icosahedral spherical
fullerenes is described by the formula $20(n^2+nl+l^2)$ with integer
$n$ and $l$. Other fullerenes are either slightly (as $C_{70}$) or
remarkably deformed and their general nickname is a "rugby ball".
The electronic structure of $C_{70}$ cluster of $D_{5h}$ geometry
has been studied in~\cite{Saito} by using the local-density
approximation in the density-functional theory. It was clearly shown
that the spheroidal geometry is of a decisive importance for
observed peculiarities in electronic states of the $C_{70}$. What is
important, the calculated energy levels were found to be in good
agreement with photoemission experiments. Spheroidal fullerenes can
be considered as a initially flat hexagonal network which has warped
into closed monosurface by twelve disclinations. We are interested
here in the slightly elliptically deformed big fullerenes like
C$_{240}$ and C$_{540}$ molecules, see
Figures~\ref{fig7},~\ref{fig8}~\cite{kroto1}.
It should be noted that the formulation of any continuum model
begins from the description of a graphite sheet (graphene) (see,
e.g.,~\cite{mele,kane,Wallace,Lomer,Weiss,Luttinger}). The models
which involve the influence of both geometry and topological defects
on the electronic structure by boundary conditions were developed in
Refs.~\cite{Ando,Dresselhaus}. A different variant of the continuum
model within the effective-mass description for fullerene near one
pentagonal defect was suggested in previous section. There are
attempts to describe electronic structure of the graphene
compositions by the Dirac-Weyl equations on the curved surfaces
where the structure of the graphite with pentagonal rings taken into
account is imposed with the help of the so-called $K$ spin
contribution~\cite{jose92,Crespi}. In the case of spherical
geometry, this is realized by introducing an effective field due to
magnetic monopole placed at the center of a sphere~\cite{jose92}. In
this section we use a similar approach. The difference is that we
consider also the elastic contribution. Indeed, pentagonal rings
being the disclination defects are the sources of additional strains
in the hexagonal lattice. Moreover, the disclinations are
topological defects and the elastic flow due to a disclination is
determined by the topological Frank index. For this reason, this
contribution exists even within the so-called "inextensional" limit
which is rather commonly used in description of graphene
compositions (see, e.g.,~\cite{Tersoff}).

Recently, in the framework of a continuum approach the exact
analytical solution for the low energy electronic states in
icosahedral spherical fullerenes has been found~\cite{Osipov}. The
case of elliptically deformed fullerenes was studied
in~\cite{Pincak} where some numerical results were presented. In
this section, we suggest a similar to~\cite{Osipov} model with the
Dirac monopole instead of 't Hooft-Polyakov monopole for describing
elastic fields. We consider a slightly elliptically deformed sphere
with the eccentricity of the spheroid $e\ll 1$. In this case, by
analogy with~\cite{Clemenger} we use the spherical representation
for the eigenstates, with the slight asphericity considered as a
perturbation. This allows us to find explicitly the low-lying
electronic spectrum for spheroidal fullerenes.

The problem of Zeeman splitting and Landau quantization of electrons
on a sphere was studied in Ref.~\cite{Aoki}. To this end, the
Schr\"odinger equation for a free electron on the surface of a
sphere in a uniform magnetic field was formulated and solved.Our
studies also cover slightly elliptically deformed fullerenes
influent by the weak uniform external magnetic field pointed in the
different directions.

\section[]{The model}

Let us start with introducing spheroidal coordinates and writing
down the Dirac operator for free massless fermions on the
Riemannian spheroid $S^{2}$. The Euler's theorem for graphene
requires the presence of twelve pentagons to get the closed
molecule. In a spirit of a continuum description we will extend
the Dirac operator by introducing the Dirac monopole field with
charge G inside the spheroid to simulate the elastic vortices due
to twelve pentagonal defects. The $K$ spin flux which describes
the exchange of two different Dirac spinors in the presence of a
conical singularity will also be included in a form of
t'Hooft-Polyakov monopole with charge g.

The Dirac equation on a surface $\Sigma$ in the presence of the
abelian magnetic monopole field $W_{\mu}$ and the external magnetic
field $A_{\mu}$ is written as~\cite{jose92,Crespi,Osipov}
\begin{equation}
i\gamma^{\alpha}e_{\alpha}^{\ \mu}[\nabla_{\mu} - iW_{\mu}-ia_{\mu}
-iA_{\mu}]\psi = E\psi. \label{eq:5.1}
\end{equation}
We impose the $K$ spin flux by the gauge field. In Cartesian
coordinates
\begin{equation}
a_{x}=\frac{2gyz}{r(r^{2}-z^{2})}\tau _{2},\
a_{y}=\frac{-2gxz}{r(r^{2}-z^{2})}\tau_{2},\ a_{z}=0.
\end{equation}
Disclination field we choose in the form
\begin{equation}
W_{x}=\frac{-G y}{r(r+z)},\ W_{y}=\frac{Gx}{r(r+z)},\ W_{z}=0,
\end{equation}
in the region $R_{N}$ (northern hemisphere) and
\begin{equation}
W_{x}=\frac{G y}{r(r-z)},\ W_{y}=\frac{-Gx}{r(r-z)},\ W_{z}=0,
\end{equation}
in the region $R_{S}$ (southern hemisphere). Here
$r=\sqrt{x^{2}+y^{2}+z^{2}}$.


\section[]{Electronic states near the Fermi energy in weak magnetic
field pointed in the z direction}

\begin{figure*}[t]
\centerline{\includegraphics[width=5.cm,clip=true]{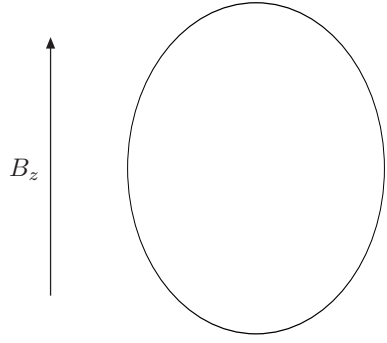}}
\caption{The schematic picture of the spheroidal fullerene in a weak
uniform magnetic field pointed in the z direction.} \label{fig9}
\end{figure*}

Firstly we will study the structure of electron levels near the
Fermi energy influent by the weak magnetic field pointed in the $z$
direction (see Fig.~\ref{fig9}) so that
$\vec{A}=B\left(y,-x,0\right)/2$.
The elliptically deformed sphere or a spheroid
\begin{equation}
\frac{x^{2}}{a^{2}}+\frac{y^{2}}{a^{2}} +\frac{z^{2}}{c^{2}}=1,
\label{eq:5.16}
\end{equation}
may be parameterized by two spherical angles $q^{1}=\phi$,
$q^{2}=\theta$ that are related to the Cartesian coordinates $x$,
$y$, $z$ as follows
\begin{equation}
x=a~\sin\theta \cos\phi; \quad y=a~\sin\theta \sin\phi; \quad
z=c~\cos\theta. \label{eq:5.17}
\end{equation}
The nonzero components of the metric tensor for spheroid are
\begin{equation}
g_{\phi\phi}=a^{2}\sin^{2}\theta; \quad
g_{\theta\theta}=a^{2}\cos^{2}\theta+c^{2}\sin^{2}\theta, \quad
\label{eq:5.18}
\end{equation}
where~$a,c\geq0,~0\leq\theta\leq\pi,~0\leq\phi<2\pi$. Accordingly,
orthonormal frame on spheroid is
\begin{equation}
e{_{1}}=\frac{1}{a~\sin\theta}~\partial_{\phi};\quad
e{_{2}}=\frac{1}{\sqrt{a^{2}\cos^{2}\theta+c^{2}\sin^{2}\theta}}~\partial_{\theta}
\label{eq:5.19}~,
\end{equation}
and dual frame reads
\begin{equation}
e{^{1}}=a~\sin\theta~d\phi;\quad
e{^{2}}=\sqrt{a^{2}\cos^{2}\theta+c^{2}\sin^{2}\theta}~d{\theta}\label{eq:5.20}
~.
\end{equation}
A general representation for zweibeins is found to be
\begin{equation}
e{^{1}}_{\phi}=a~\sin\theta;\  e^{1}_{\ \theta}=0;\ e^{2}_{\
\phi}=0;\
e{^{2}}_{\theta}=\sqrt{a^{2}\cos^{2}\theta+c^{2}\sin^{2}\theta}
\label{eq:5.21} ~.
\end{equation}
Notice that $e^{\mu}_{\ \alpha}$ is the inverse of $e^{\alpha}_{\
\mu}$. The Riemannian connection with respect to the orthonormal
frame is written as~\cite{Nakahara,Gockeler}
\begin{equation}
de^{1}=-\omega^{1}_{\ 2}\wedge e^{2}=\frac{a
\cos\theta}{\sqrt{a^{2}\cos^{2}\theta+c^{2}\sin^{2}\theta}}\
d\phi\wedge e^{2}, \label{eq:5.22}
\end{equation}
\begin{equation}
de^{2}=-\omega^{2}_{\ 1}\wedge e^{1}=0. \label{eq:5.23}
\end{equation}
Here $\wedge $ denotes the exterior product and $d$ is the exterior
derivative. From Eqs.~\ref{eq:5.22} and ~\ref{eq:5.23} we get the
Riemannian connection in the form
\begin{equation}
\omega{_{\phi 2}^{1}}=-\omega{_{\phi 1}^{2}}=\frac{a
\cos\theta}{\sqrt{a^{2}\cos^{2}\theta+c^{2}\sin^{2}\theta}}\ ; \quad
\omega{_{\theta 2}^{1}}=\omega{_{\theta 1}^{2}}=0. \label{eq:5.24}
\end{equation}
We assume that the eccentricity of the spheroid
$e=\sqrt{|1-(c/a)^{2}|}$ is small enough. In this case, one can
write down $c=a+\delta a$ where $\delta$ ($\mid\delta\mid\ll1$) is a
small dimensionless parameter characterizing the spheroidal
deformation from the sphere. So, we can follow the perturbation
scheme using $\delta$ as the perturbation parameter. Within the
framework of the perturbation scheme the spin connection
coefficients are written as
\begin{equation}
\omega{_{\phi 2}^{1}}=-\omega{_{\phi
1}^{2}}\approx\cos\theta(1-\delta\sin^{2}\theta), \label{eq:5.25}
\end{equation}
where the terms to first order in $\delta$ are taken into account.
The inclusion of the K-spin connection can be performed by
considering two Dirac spinors as the two components of an $SU(2)$
color doublet $\psi= (\psi_\uparrow \psi_\downarrow)^T$ (see
Ref.~\cite{jose92} for details). In this case, the interaction with
the color magnetic fields in a form of nonabelian magnetic flux ('t
Hooft'-Polyakov monopole) is responsible for the exchange of the two
Dirac spinors which in turn corresponds to the interchange of Fermi
points. As was shown in Ref.~\cite{jose91}, the Dirac equation for
$\psi$ can be reduced to two decoupled equations for $\psi_\uparrow$
and $\psi_\downarrow$, which include an Abelian monopole of opposite
charge, $g=\pm 3/2$.

The elastic flow through a surface due to a disclination has a
vortex-like structure~\cite{jetplet00} and can be described by
Abelian gauge field $W_\mu$. Similarly to Ref.~\cite{jose92}, we
replace the fields of twelve disclinations by the effective field of
the magnetic monopole of charge $G$ located at the center of the
spheroid.
The circulation of this field is determined by the Frank index,
which is the topological characteristic of a disclination. Trying to
avoid an extension of the group we will consider the case of the
Dirac monopole. The another possibility is to introduce the
non-Abelian 't Hooft-Polyakov monopole (see Ref.~\cite{Osipov}). As
is known, the vector potential $W_\mu$ around the Dirac monopole
have singularities. To escape introducing singularities in the
coordinate system let us divide the spheroid (similarly as it was
done for the sphere in~\cite{Yang}) into two regions, $R_{N}$ and
$R_{S}$, and define a vector potential $W_{\mu}^{N}$ in $R_{N}$ and
a vector potential $W_{\mu}^{S}$ in $R_{S}$. Notice that below
$W_\mu$ includes both gauge fields $W_\mu$ and $a_\mu$ defined in
Eq.~\ref{eq:5.1}.

One has
$$
R_{N}:\ \ 0\leq\theta<\frac{\pi}{2}+\Delta;\ 0\leq\phi<2\pi,
$$
$$
R_{S}:\ \ \frac{\pi}{2}-\Delta<\theta\leq\pi; \ 0\leq\phi<2\pi,
$$
\begin{equation}
R_{NS}:\ \ \frac{\pi}{2}-\Delta<\theta<\frac{\pi}{2}+\Delta;\
0\leq\phi<2\pi, \ (overlap) \label{eq:5.26}
\end{equation} where
$\Delta$ is chosen from the interval $0<\Delta\leq\pi/2$. In
spheroidal coordinates~(\ref{eq:5.18}) the only nonzero components
of $W_{\mu}$ are found to be
\begin{equation}
W^{N}_{\phi}\approx g\cos\theta(1+\delta\sin^{2}\theta) +
G(1-\cos\theta)-\delta G\sin^{2}\theta
\cos\theta,\label{eq:5.27}\end{equation}
\begin{equation}
 W^{S}_{\phi}\approx g\cos\theta(1+\delta\sin^{2}\theta)
 -G(1+\cos\theta)-\delta G\sin^{2}\theta \cos\theta,
\label{eq:5.28}
\end{equation}
where the terms of an order of $\delta^2$ and higher are dropped. In
the overlapping region the potentials $W_{\phi}^N$ and $W_{\phi}^S$
are connected by the gauge transformation
\begin{equation}
W_{\phi}^{N}=W_{\phi}^{S}+iS_{NS}\ \partial_{\phi}S^{-1}_{NS},
\label{eq:5.29}
\end{equation}
where
$$S_{NS}=e^{2iG\phi}$$
is the phase factor. The wavefunctions in the overlap $R_{NS}$ are
connected by
\begin{equation}
\psi_{N}=S_{NS}\psi_{S}, \label{eq:5.30}
\end{equation}
where $\psi_{N}$ and $\psi_{S}$ are the spinors in $R_{N}$ and
$R_{S}$, respectively. Since $S_{NS}$ must be
single-valued~\cite{Wu}, $2G$ takes integer values. Notice that the
total flux due to Dirac monopole in Egs.~\ref{eq:5.27} and
~\ref{eq:5.28} is equal to $4\pi G$. What is important, there is no
contribution from  the terms with $\delta$. In our case, the total
flux describes a sum of elastic fluxes due to twelve disclinations,
so that the total flux (the modulus of the total Frank vector) is
equal to $12\times \pi/3=4\pi$. Therefore, one obtains $G=1$.

Let us consider the region $R_{N}$. In spheroidal coordinates, the
only nonzero component of $W_{\mu}$ in region $R_{N}$ is found to be
\begin{equation}
W_{\phi}\approx g\cos\theta(1+\delta\sin^{2}\theta) +
G(1-\cos\theta)-\delta G\sin^{2}\theta
\cos\theta,\label{eq:5.31}\end{equation} and the external magnetic
field pointed in the z-direction reads
\begin{equation}
A_{\phi}=-\frac{1}{2}Ba^{2}\sin^{2}\theta.\label{eq:5.32}
\end{equation}
By using the substitution (\ref{eq:5.8}) we obtain the Dirac
equation for functions $u_j$ and $v_j$ in the form
$$
\left(-i\sigma_{1}\frac{1}{a}(\partial_{\theta}+\frac{\cot\theta}{2})+\frac{\sigma_{2}}{a
\sin\theta}\left(j-m\cos\theta+\frac{1}{2}Ba^{2}\sin^{2}\theta\right)+\delta
\hat{\cal{D}}_{1}\right) \left(\begin{array}{c}
  u_j(\theta) \\
  v_j(\theta) \\
\end{array}
\right)\nonumber
$$
\begin{equation}
=E\left(\begin{array}{c}
  u_j(\theta) \\
  v_j(\theta) \\
\end{array}
\right), \label{eq:5.33}
\end{equation}
where
\begin{equation}
\hat{\cal{D}}_{1}=-\frac{\gamma_{1}}{a}\sin\theta\left(j-2m\cos\theta\right)-\gamma_{1}\frac{B
a}{2}\sin^{3}\theta. \label{eq:5.34}
\end{equation}
A convenient way to study the eigenvalue problem is to square Eq.
(\ref{eq:5.33}). For this purpose, let us write the Dirac operator
in Eq. (\ref{eq:5.33}) as
$\hat{\cal{D}}=\hat{\cal{D}}_{0}+\delta\hat{\cal{D}}_{1}$. One can
easily obtain that
\begin{eqnarray}
\hat{\cal{D}}_{0}^2&=&-\frac{1}{a^{2}}\left(\partial^{2}_{\theta}+\frac{\cos\theta}{\sin\theta}\partial_{\theta}-\frac{1}{4}-\frac{1}{4
\sin^{2}\theta}\right)+\frac{\left(j-m\cos\theta\right)^{2}}{a^{2}\sin^{2}\theta}+Bj
+\frac{B}{2}\left(\sigma_{3}-2m\right)\cos\theta \nonumber\\
&&+\sigma_{3}\frac{m-j\cos\theta}{a^{2}\sin^{2}\theta}+\frac{B^{2}a^{2}\sin^{2}\theta}{4}.
\end{eqnarray}
Notice that to first order in $\delta$ the square of the Dirac
operator is written as $
\hat{\cal{D}}^{2}=(\hat{\cal{D}}^{2}_{0}+\delta\hat\Gamma), $ where
$
\hat\Gamma=(\hat{\cal{D}}_{0}\hat{\cal{D}}_{1}+\hat{\cal{D}}_{1}\hat{\cal{D}}_{0}).$
In an explicit form
$$
a^{2}\hat\Gamma=2j^{2}+jx\left(\sigma_{3}-6m\right)
+4m(m-\sigma_{3})x^{2}+2m\sigma_{3}+3Ba^{2}(1-x^{2})x(\sigma_{3}/2-m)
$$
\begin{equation}
 +2Bja^{2}(1-x^{2})+B^{2}a^{4}(1-x^{2})^{2}/2,
\label{eq:5.35}
 \end{equation}
where the appropriate substitution $x=\cos\theta$ is used. The
equation $\hat{\cal{D}}^2\psi=E^2\psi$ takes the form
$$
[\partial_x(1-x^2)\partial_x-\frac{(j-mx)^2-j\sigma_3
x+\frac{1}{4}+\sigma_3 m}{1-x^2}-a^{2}B V(x) -\delta a^{2}\hat\Gamma]\left(%
\begin{array}{c}
  u_j(x) \\
  v_j(x) \\
\end{array}%
\right)\nonumber
$$
\begin{equation}
=-(\lambda^2-\frac{1}{4})\left(%
\begin{array}{c}
  u_j(x) \\
  v_j(x) \\
\end{array}%
\right), \label{eq:5.36}
\end{equation}
where $\lambda=aE, V(x)=j+(\sigma_{3}-2m)x/2$. Since we consider the
case of a weak magnetic field the terms with $B^{2}$ and $\delta B$
in Eq. (\ref{eq:5.36}) can be neglected. As a result, the energy
spectrum for spheroidal fullerenes is found to be

\begin{equation}
(\lambda_{jn}^{\delta})^2=(n+|j|+1/2)^2-m^2+Ba^{2}\left(j
+A_{jn}\right)+\delta\big(2j^{2}+A^{1}+A^{2}\big) \label{eq:5.37}
\end{equation}
where $$ A_{jn}=-\frac{j(m^2-1/4)}{p(p+1)},$$ and
\begin{equation}
A^{1}=-\frac{j^{2}(6m^{2}-1/2)}{p(p+1)}+\frac{j^{2}m^{2}(4m^{2}-3)}{p^{2}(p+1)^{2}},\label{eq:5.38}
\end{equation}
\begin{equation}
A^{2}=\frac{\left(4m(m-1)F_{n}(2\alpha,2\beta)-X\right)T\
J_{n}(2\alpha,2\beta)
+\left(4m(m+1)F_{n}(2\mu,2\nu)+X\right)J_{n}(2\mu,2\nu)}{T\
J_{n}(2\alpha,2\beta)+J_{n}(2\mu,2\nu)}\label{eq:5.39}
\end{equation}
\begin{equation}
 X=\frac{2j^{2}m}{p(p+1)}+\frac{j^{2}m }{p^2(p+1)^{2}}-2m\label{eq:5.40}
\end{equation}
\begin{equation}
F_{n}(2\alpha,2\beta)=\frac{(n+1)(n+2\alpha+1)(n+2\beta+1)(n+2\alpha+2\beta+1)}
{(2p+1)(p+1)^{2}(2p+3)}\label{eq:5.41}\end{equation}
$$
+\frac{n(n+2\alpha)(n+2\beta)(n+2\alpha+2\beta)}{(2p-1)p^{2}(2p+1)}
$$
$$
J_{n}(2\alpha,2\beta)=\Gamma(n+2\alpha+1) \Gamma(n+2\beta+1),
$$
Here $p=n+\beta+\alpha$ and
\begin{equation}
T=\frac{n+\mid j\mid-m+1/2}{n+\mid j\mid+m+1/2}\label{eq:5.42}
\end{equation}

When $\delta=0, B=0$, one has the case of a sphere with magnetic
monopole inside. In this case, the exact solution is known (see,
e.g.,~\cite{Osipov})
\begin{eqnarray}
\left[\partial_x(1-x^2)\partial_x-\frac{(j-mx)^2-j\sigma_3
x+\frac{1}{4}+\sigma_3 m}{1-x^2}\right]\left(%
\begin{array}{c}
  u_{jn}^{0}(x) \\
  v_{jn}^{0}(x) \\
\end{array}%
\right)=-(\lambda^2_{0n}-\frac{1}{4})\left(%
\begin{array}{c}
  u_{jn}^{0}(x) \\
  v_{jn}^{0}(x) \\
\end{array}
\right), \label{eq:27}\nonumber\end{eqnarray}

\begin{eqnarray}
\left(%
\begin{array}{c}
  u_{jn}^{0} \\
  v_{jn}^{0} \\
\end{array}%
\right)
=\left(%
\begin{array}{c}
  C_{u}(1-x)^{\alpha}(1+x)^{\beta}P_{n}^{2\alpha,2\beta} (x) \\
  C_{v}(1-x)^{\mu}(1+x)^{\nu}P_{n}^{2\mu,2\nu} (x) \\
\end{array}%
\right),
\end{eqnarray}
with the energy spectrum
\begin{equation}
(\lambda_{jn}^{0})^2=(n+|j|+1/2)^2-m^2. \label{eq:28}
\end{equation}
Here
\begin{eqnarray}
\alpha=\frac{1}{2}\left|j-m-\frac{1}{2}\right|,\beta=\frac{1}{2}\left|j+m+\frac{1}{2}\right|,\nonumber\\
\mu=\frac{1}{2}\left|j-m+\frac{1}{2}\right|,\nu=\frac{1}{2}\left|j+m-\frac{1}{2}\right|,
\label{eq:29}
\end{eqnarray}
$P_{n}^{2\alpha,2\beta} (x)$ and $P_{n}^{2\mu,2\nu} (x)$ are Jacobi
polynomials of $n$-th order, and $C_{u}$ and $C_{v}$ are the
normalization factors. These states are degenerate. For example, the
degeneracy of the zero mode is equal to six. Therefore, we have to
use the perturbation scheme for the degenerate energy
levels~\cite{Landau}.

Finally, in the linear in $\delta$ approximation, the low energy
electronic spectrum of spheroidal fullerenes takes the form
\begin{equation}
E_{jn}=E^0_{jn}+E^{\delta B_{z} }_{jn} \label{eq:5.44}
\end{equation}
with
\begin{eqnarray}
E^0_{jn}=\pm\sqrt{(2\xi+n)(2\eta+n)}, \nonumber\\
E^{\delta B_{z} }_{jn}=\frac{Ba^{2}(j+A_{jn})+\delta
(2j^{2}+A^{1}+A^{2})}{2E^0_{jn}}, \label{eq:5.45}
\end{eqnarray}
where $\xi=\mu\ (\nu)$ and $\eta=\beta\ (\alpha)$ for $j>0\  (j<0)$,
respectively. The parameters $\alpha$, $\beta$, $\mu$, $\nu$ are
defined above.

Here we came back to the energy variable $E=\lambda/a$ (in units of
$\hbar V_F/a$ where $V_F$ is the Fermi velocity). Notice that the
non-diagonal matrix element of perturbation $\langle
e^{j\phi}|\langle\psi_{jn}|\Gamma|\psi_{-jn}\rangle|e^{-j\phi}\rangle$
turns out to be zero. Therefore, the states with opposite $j$ do not
mix and $j$ remains a good quantum number as would be expected. As
is seen from Eq. (\ref{eq:5.44}), for $\delta=0$ both eigenstates
and eingenvalues are the same as for a sphere (cf.
Ref.~\cite{Osipov}). At the same time, Eqs. (\ref{eq:5.44}) and
(\ref{eq:5.45}) show that the spheroidal deformation gives rise to
an appearance of fine structure in the energy spectrum. The
difference in energy between sublevels is found to be linear in
$\delta$ which resembles the Zeeman effect where the splitting
energy is linear in magnetic field. In addition we found a further
splitting of the states with opposite $j$.
\begin{table}[htb]
\begin{center}
\begin{tabular}[textwidth]{l c c c c} \bfseries\bfseries $YO-C_{240}$\quad & $j$ &
\bfseries $E^{0}_{jn} (eV)$
&\bfseries $E^{\delta }_{jn}(meV)(B=0)$& \bfseries $ E^{\delta B_z}_{jn} (meV)(B a^{2}=0.1)$\\
\hline\hline\bfseries $n=0$, $m=1/2$
&1&1.094&10.5&37.5\\
&-1&1.094&10.5&-16.5\\
\hline\hline \bfseries $n=1$, $m=1/2$
&1&1.89&8.8&24.8\\
&-1&1.89&8.8&-7.2\\
\bfseries  $n=0$, $m=1/2$
&2&1.89&28.4&60.4\\
&-2&1.89&28.4&-3.6\\
\bfseries $n=0$, $m=-5/2$
&3&1.89&3&27\\
&-3&1.89&3&-21\\
\end{tabular}
\end{center}
\caption{{\footnotesize The structure of the first and higher energy
levels for YO-C$_{240}$ fullerene in uniform magnetic field. The
hopping integral and other parameters are taken to be $t=2.5\ eV$
and $V_F=3t\overline{b}/2\hbar$, $\overline{b}=1.45{\textmd{\AA}}$,
$\overline{R}=7.03{\textmd{\AA}}$, $SD=0.17{\textmd{\AA}}$,
$\delta=0.024$. $\overline{b}$ is average bond length,
$\overline{R}\ (\overline{R}=a)$ is average radius, $SD$ is standard
deviation from a perfect sphere (see Refs.~\cite{yoshida,Lu}), so
that $\delta=SD/\overline{R}$.}} \label{tab1}
\end{table}

\begin{figure*}[htb]
\centerline{\includegraphics[width=6.cm,clip=true]{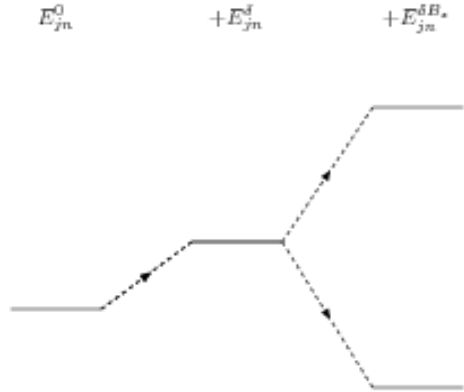}}
\caption{The schematic picture of the first positive electronic
level $E_{jn}^{\delta}$ for spheroidal fullerenes in a weak uniform
magnetic field pointed in the z direction.}\label{fig10}
\end{figure*}

\begin{figure*}[htb]
\centerline{\includegraphics[width=6.cm,clip=true]{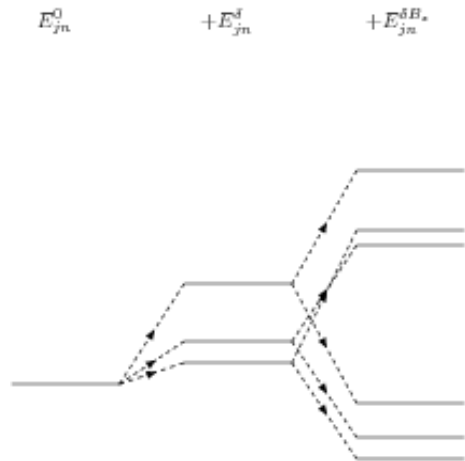}}
\caption{The schematic picture of the second positive electronic
level $E_{jn}^{\delta}$ of the spheroidal fullerenes in a weak
uniform magnetic field pointed in the z direction.}\label{fig11}
\end{figure*}

As an example, Table~\ref{tab1} shows all contributions (in
compliance with Eq. (\ref{eq:5.44})) to the first and second energy
levels for YO-C$_{240}$ (YO means a structure given in
\cite{yoshida,Lu}). Schematically, the structure of the first energy
level is shown in Fig.~\ref{fig10}. As is clearly seen, the first
(double degenerate) level becomes shifted due to spheroidal
deformation and splitted due to nontrivial topology. The uniform
magnetic field provides the well-known Zeeman splitting. The
difference between topological and Zeeman splitting is clearly seen.
In the second case, the splitted levels are shifted in opposite
directions while for topological splitting the shift is always
positive. The more delicate is the structure of the second level
which is schematically presented in Fig.~\ref{fig11}. In this case,
the initial (for $\delta=0$) degeneracy of $E^0_{jn}$ is equal to
six. The spheroidal deformation provokes an appearance of three
shifted double degenerate levels (fine structure) which, in turn,
are splitted due to the presence of topological defects. The
magnetic field is responsible to Zeeman splitting.

We admit that the obtained values of the splitting energies can
deviate from the estimations within some more precise microscopic
lattice models and density-functional methods (see,
e.g.\cite{Broglia}). Therefore, we focused mostly on the very
existence of physically interesting effects. For this reason, the
values in both Table and the schematic pictures are presented with
accuracy at about one percent of $E^{0}_{jn}$.

\newpage

\section[]{Electronic states near the Fermi energy in weak magnetic field pointed
in the x direction}

\begin{figure*}[htb]
\centerline{\includegraphics[width=5.cm,clip=true]{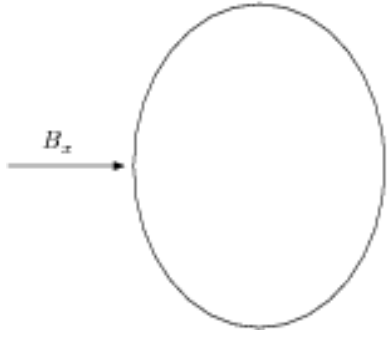}}
\caption{The schematic picture of the spheroidal fullerene in a weak
uniform magnetic field pointed in the x direction.}\label{fig12}
\end{figure*}

In the second case we introduce that the uniform external magnetic
field $B$ is chosen to be pointed in the $x$ direction
 (see~\ref{fig12}) so that $\vec{A}=B\left(0,-z,y\right)/2$.

The nonzero component of $W_{\mu}$ in region $R_{N}$ are the same as
in previous case but the external magnetic field are changed
\begin{equation}
A_{\phi}=-\frac{1}{2}Bac\sin\theta\cos\theta\cos\phi,\label{eq:5.47}
\end{equation}
\begin{equation}
A_{\theta}=-\frac{1}{2}Bac\sin\phi,\label{eq:5.48}
\end{equation}
where $c=a+\delta a$ and only the terms to the first order in
$\delta$ are taken into account. Following the calculations above
with neglecting the terms $B^2$ and $\delta B$ we obtain the Dirac
equation for functions $u_j$ and $v_j$ in the form
$$
\left(-i\sigma_{1}\frac{1}{a}(\partial_{\theta}+\frac{\cot\theta}{2}+i\frac{1}{2}Ba^{2}\sin\phi)+\frac{\sigma_{2}}{a
\sin\theta}\left(j-m\cos\theta+\frac{1}{2}Ba^{2}\sin\theta\cos\theta\cos\phi\right)+\delta
\hat{\cal{D}}_{1}\right) \times
$$
\begin{equation}
\times\left(\begin{array}{c}
  u_j(\theta) \\
  v_j(\theta) \\
\end{array}
\right)=E\left(\begin{array}{c}
  u_j(\theta) \\
  v_j(\theta) \\
\end{array}
\right), \label{eq:5.49}
\end{equation}
where
\begin{equation}
\hat{\cal{D}}_{1}=-\frac{\gamma_{1}}{a}\sin\theta\left(j-2m\cos\theta\right)-\gamma_{1}\frac{B
a}{2}\sin^{2}\theta\cos\theta\cos\phi. \label{eq:5.50}
\end{equation}
We find the square of nonperturbative part of Dirac operator above
in the form
\begin{eqnarray}
\hat{\cal{D}}_{0}^2&=&-\frac{1}{a^{2}}\left(\partial^{2}_{\theta}+\frac{\cos\theta}{\sin\theta}\partial_{\theta}-\frac{1}{4}-\frac{1}{4
\sin^{2}\theta}\right)+\frac{\left(j-m\cos\theta\right)^{2}}{a^{2}\sin^{2}\theta}\nonumber\\
&&+\sigma_{3}\frac{m-j\cos\theta}{a^{2}\sin^{2}\theta}+BV(\theta,\phi),
\label{eq:5.51}
\end{eqnarray}
where the terms with $B^{2}$ and $\delta B$ was neglecting and
\begin{eqnarray}
V(\theta,\phi)&=&-i(\cos\phi\partial_\phi-\frac{\sin\phi}{2})\cot\theta-\sigma_3\frac{\cos\phi(1+\sin^{2}\theta)}{2\sin\theta}-i(\partial_\theta+\frac{1}{2}\cot\theta)\sin\phi
\nonumber
\\
&&-\frac{m \cos^{2}\theta\cos\phi}{\sin\theta}.
\end{eqnarray}
We denote
$V_{j,j'}^{m}=\langle\psi_{jn}^{m}|V(\theta,\phi)|\psi_{j'n}^{m}\rangle$,
where $\psi_{jn}^{m} $ are the eigenfunctions of $\hat{\cal{D}}_{0}
$. In the case of first energy mode the terms $V_{-1,-1}^{1/2}$,
$V_{1,1}^{1/2}$, $V_{1,-1}^{1/2}$, $V_{-1,1}^{1/2} $ are zero and
therefore the weak magnetic field pointed in the x direction do not
influent the first energy level. For the second energy mode only
nondiagonal terms $V_{-1,-2}^{1/2}$, $V_{1,2}^{1/2}$ were find
different from zero. The following numerical values of the
nondiagonal terms $V_{j,j'}^{m}$ was found: $V_{-1,-2}^{1/2}=-2.1$,
$V_{1,2}^{1/2}=2.1$. We use the wave functions for calculations of
nondiagonal terms in the case of second energy level in the form:
\begin{eqnarray}
\psi_{0,2}^{1/2} (z,\phi)= \frac{e^{i2\phi}}{\sqrt{2\pi}}\sqrt{\frac{32}{15}}\left(%
\begin{array}{c}
  -i\sqrt{\frac{2}{3}}\sqrt{1-z^{2}}(1+z)\\
   1-z^{2}\\
\end{array}%
\right) ,  \nonumber \\
\psi_{0,-2}^{1/2} (z,\phi)= \frac{e^{-i2\phi}}{\sqrt{2\pi}}\sqrt{\frac{32}{15}}\left(%
\begin{array}{c}
  i\sqrt{\frac{2}{3}}\sqrt{1-z^{2}}(1-z)\\
   1-z^{2}\\
\end{array}%
\right) ,\label{eq:5.52}
\end{eqnarray}

\begin{eqnarray}
\psi_{1,1}^{1/2} (z,\phi)= \frac{e^{i\phi}}{\sqrt{2\pi}}\sqrt{\frac{3}{5}}\left(%
\begin{array}{c}
  -i\sqrt{\frac{2}{3}}(2z^{2}+z-1)\\
   \frac{3}{2}z\sqrt{1-z^{2}}\\
\end{array}%
\right) ,  \nonumber\\
\psi_{1,-1}^{1/2} (z,\phi)= \frac{e^{-i\phi}}{\sqrt{2\pi}}\sqrt{\frac{3}{5}}\left(%
\begin{array}{c}
  i\sqrt{\frac{2}{3}}(-2z^{2}+z+1)\\
   \frac{3}{2}z\sqrt{1-z^{2}}\\
\end{array}%
\right) .\label{eq:5.53}
\end{eqnarray}

The low energy electronic spectrum of spheroidal fullerenes in this
case  takes the form
\begin{equation}
E_{jn}=E^0_{jn}+E^{\delta B_{x}},\label{eq:5.54}
\end{equation}
where
\begin{equation}
E^{\delta B_{x}}=\frac{\delta
\left(\hat{\Gamma}_{22}+\hat{\Gamma}_{11}\right)\pm\sqrt{\delta^{2}
\left(\hat{\Gamma}_{22}-\hat{\Gamma}_{11}\right)^{2}+4|Ba^{2}V_{1,2}|^2}}{4
E^0_{jn}}\label{eq:5.55} .\end{equation}

\begin{figure*}[htb]
\centerline{\includegraphics[width=7.cm,clip=true]{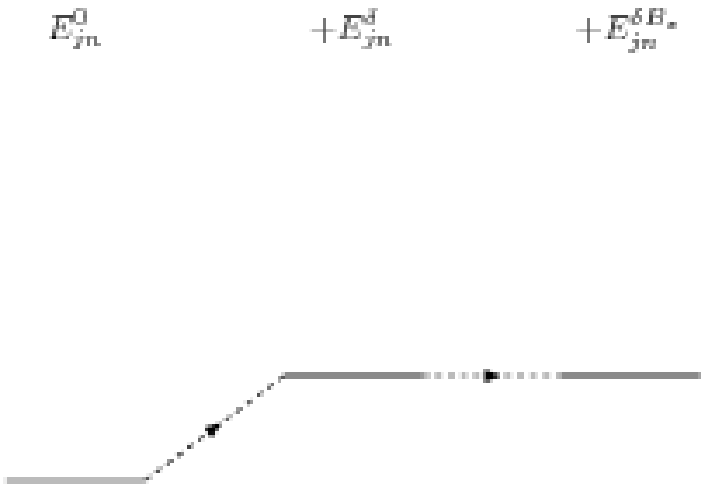}}
\caption{The schematic picture of the first positive electronic
level $E_{jn}^{\delta}$ for spheroidal fullerenes in a weak uniform
magnetic field pointed in the x direction.}\label{fig13}
\end{figure*}

\begin{figure*}[htb]
\centerline{\includegraphics[width=7.cm,clip=true]{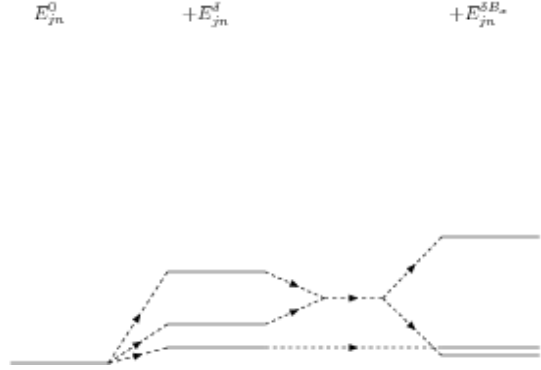}}
\caption{The schematic picture of the second positive electronic
level $E_{jn}^{\delta}$ with a mixture of quantum numbers $j=1,2$
and $j=-1,-2$ for spheroidal fullerenes in a weak uniform magnetic
field pointed in the x direction.}\label{fig14}
\end{figure*}

We change $1,2$ to $-1,-2$ in the expression above to get the
another energy levels in the degenerate energy state which is
considered. Where the $\hat{\Gamma}_{ii}$ are the diagonal matrix
element of perturbation of Eq.~\ref{eq:5.35} with $B=0$.
Table~\ref{tab2} shows all contributions to the first and second
energy levels for YO-C$_{240}$ fullerenes influent by the horizontal
weak external magnetic field. Figures~\ref{fig13},~\ref{fig14}
present first and second positive energy levels of spheroidal
fullerenes in weak magnetic field pointed in the x direction in
comparison with Figures~\ref{fig10},~\ref{fig11}. As is seen, there
is a marked difference between the behavior of the first and second
energy levels in magnetic field. Indeed, in both cases the energy
levels become shifted due to spheroidal deformation. However, the
uniform magnetic field does not influence the first energy level.
The splitting takes place only for the second level. We can conclude
that there is a possibility to change the structure of the
electronic levels in spheroidal fullerenes by altering the direction
of magnetic field. It would be interesting to test this prediction
in experiment. The values in both Table and the schematic pictures
are presented the same way as in the previous case with accuracy at
about one percent of $E^{0}_{jn}$.

\begin{table}[htb]
\begin{center}
\begin{tabular}{l c c c c c }
\bfseries\bfseries $YO-C_{240}$\quad \quad & $j$ & \bfseries
$E^{0}_{jn} (eV)$ & \bfseries $ E_{jn}^{\delta }(meV)(B=0)$
& \bfseries $E_{jn}^{\delta B_{x}}(meV)(B a^{2}=0.1)$\\
\hline\hline\bfseries $n=0$, $m=1/2$
&1&1.094&10.5&10.5\\
&-1&1.094&10.5&10.5\\
\hline\hline\bfseries $n=0$, $m=-5/2$
&3&1.89&3&3\\
&-3&1.89&3&3\\
\hline\bfseries  $n=0$, $m=1/2$
&2&1.89&28.4&\\
&&&& 53/-16/\\
$n=1$&1&1.89&8.8&\\
\hline\bfseries  $n=1$, $m=1/2$
&-1&1.89&8.8&\\
&&&& 53/-16/\\
$n=0$&-2&1.89&28.4&\\
\end{tabular}
\end{center}
\caption{{\footnotesize The structure of the first and second energy
levels for YO-C$_{240}$ fullerene in uniform magnetic field. The
hopping integral and other parameters are taken to be $t=2.5\ eV$
and $V_F=3t\overline{b}/2\hbar$, $\overline{b}=1.45{\textmd{\AA}}$,
$\overline{R}=7.03{\textmd{\AA}}$, $SD=0.17{\textmd{\AA}}$,
$\delta=0.024$. $\overline{b}$ is average bond length,
$\overline{R}\ (\overline{R}=a)$ is average radius, $SD$ is standard
deviation from a perfect sphere (see Refs.~\cite{yoshida,Lu}), so
that $\delta=SD/\overline{R}$.}} \label{tab2}
\end{table}
\newpage
\newpage
\section[]{Zero-energy mode}

Now we will analyze the special case of an electron state at the
Fermi level (so-called zero-energy mode), in this case the right
side of equation~\ref{eq:5.1} is equal to zero. We will use the
projection coordinates in the form
\begin{equation}
x=\frac{2R^{2}r}{R^{2}+r^{2}} \cos\phi; \quad
y=\frac{2R^{2}r}{R^{2}+r^{2}} \sin\phi; \quad
z=-c\frac{R^{2}-r^{2}}{R^{2}+r^{2}},\label{eq:5.3}
\end{equation}
where $R$ and $c$ are the spheroidal axles. The Riemannian
connection with respect to the orthonormal frame is written
as~\cite{Nakahara,Gockeler}
\begin{equation}
-\omega{_{\phi 2}^{1}}=\omega{_{\phi
1}^{2}}=\frac{R^{2}-r^{2}}{\sqrt{(R^{2}-r^{2})^{2}+4c^{2}r^{2}}}\ ;
\quad \omega{_{r 2}^{1}}=\omega{_{r 1}^{2}}=0. \label{eq:5.4}
\end{equation}

The only nonzero component of the gauge field $W_{\mu}$ in region
$R_{N}$  for spheroidal fullerenes reads
\begin{equation}
W_{\phi}=G+m\frac{z}{X},\label{eq:5.5}
\end{equation}
where
\begin{equation}
X=\frac{\sqrt{c^{2}(r^2-R^2)^2+4R^{4}r^2}}{r^2+R^2}. \label{eq:5.6}
\end{equation}
Notice that the monopole field $W_{\mu}$ in Eq. (\ref{eq:5.5})
consists of two parts. The first one comes from the $K$-spin
connection term and implies the charge $g=\pm 3/2$ while the second
one is due to elastic flow through a surface. This contribution is
topological in its origin and characterized by charge $G$. For total
elastic flux from twelve pentagonal defects one has $G=1$. The
parameter $m=g-G$ is introduced in Eq.(\ref{eq:5.5}). The uniform
external magnetic field $B$ for simplicity is chosen to be pointed
in the $z$ direction. In projection coordinates the only nonzero
component of $A_\mu$ is written as
\begin{equation}
A_{\phi}=-\frac{2BR^{4}r^{2}}{(R^{2}+r^{2})^{2}}.\label{eq:5.7}
\end{equation}
Let us study Eq.(\ref{eq:5.1}) for the electronic states at the
Fermi energy ($E=0$). The Dirac matrices can be chosen to be the
Pauli matrices, $\gamma_1=-\sigma_2, \gamma_2=-\sigma_1$. By using
the substitution
\begin{equation}
\left(%
\begin{array}{c}
  \psi_{A} \\
  \psi_B \\
\end{array}%
\right) =\sum_j \frac{e^{i(j+G)\phi}}{\sqrt{2\pi}}\left(%
\begin{array}{c}
  u_j(r) \\
  v_j(r) \\
\end{array}%
\right) ,j=0,\pm 1,\pm 2,\ldots \label{eq:5.8}
\end{equation}
we obtain
\begin{equation}
\left(\frac{1}{r}(j-m\frac{z}{X}-A_{\phi})+\frac{R^2+r^2}{K}(\partial_{r}-\frac{r^2-R^2}{2r(r^2+R^2)})\right)v(r)=0,
\label{eq:5.9}
\end{equation}
\begin{equation}
\left(\frac{1}{r}(j-m\frac{z}{X}-A_{\phi})-\frac{R^2+r^2}{K}(\partial_{r}-\frac{r^2-R^2}{2r(r^2+R^2)})\right)u(r)=0,
\label{eq:5.10}
\end{equation}
where $K=\sqrt{(R^2-r^2)^2+4c^{2}r^2}$. We assume that the
eccentricity of the spheroid is small enough. In this case, one can
write down $c=R+\delta R$ and $\delta$ ($\mid\delta\mid\ll1$) is a
small dimensionless parameter characterizing the spheroidal
deformation. Therefore, one can follow the perturbation scheme using
$\delta$ as the perturbation parameter. To the leading in $\delta$
approximation Eqs. (\ref{eq:5.9}) and (\ref{eq:5.10}) are written as
$$\partial_{r}v_{j}(r)=(-\frac{(R^2+r^2)^2}{r[(R^2+r^2)^2-4R^{2}r^{2}\delta]}[j-m(1+\delta)\frac{r^2-R^2}{r^2+R^2}+
m\delta\frac{(r^2-R^2)^{3}}{(r^2+R^2)^{3}}]$$
\begin{equation}
-\frac{2BR^{4}r}{[(R^2+r^2)^2-4R^{2}r^{2}\delta]}+\frac{r^2-R^2}{2r(R^2+r^2)})v_{j}(r),
\label{eq:5.11}
\end{equation}
$$\partial_{r}u_{j}(r)=(\frac{(R^2+r^2)^2}{r[(R^2+r^2)^2-4R^{2}r^{2}\delta]}[j-m(1+\delta)\frac{r^2-R^2}{r^2+R^2}
+m\delta\frac{(r^2-R^2)^{3}}{(r^2+R^2)^{3}}]$$
\begin{equation}
+\frac{2BR^{4}r}{[(R^2+r^2)^2-4R^{2}r^{2}\delta]}+\frac{r^2-R^2}{2r(R^2+r^2)})u_{j}(r).
\label{eq:5.12}
\end{equation}
The exact solution to Eqs. (\ref{eq:5.11}) and (\ref{eq:5.12}) is
found to be
\begin{equation}
v_{j}(x)=\frac{(x+1)^{(1/2-m)}}{x^{\frac{1}{2}(j+m+1/2)}}
\left(\frac{x+1-2\delta-2\sqrt{\delta(\delta-1)}}
{x+1-2\delta+2\sqrt{\delta(\delta-1)}}\right)^{-\alpha}\left((x+1)^{2}-4\delta
x\right)^{m}, \label{eq:5.13}
\end{equation}
\begin{equation}
u_{j}(x)=(x+1)^{(m+1/2)} x^{\frac{1}{2}(j+m-1/2)}
\left(\frac{x+1-2\delta-2\sqrt{\delta(\delta-1)}}
{x+1-2\delta+2\sqrt{\delta(\delta-1)}}\right)^{\alpha}\left((x+1)^{2}-4\delta
x\right)^{-m}, \label{eq:5.14}
\end{equation}
where $x=r^2/R^2 $,
$\alpha=(j/2)\sqrt{\delta/(\delta-1)}+BR^{2}/4\sqrt{\delta(\delta-1)}
$. The function $u_{j}$ can be normalized if the condition
$m-1/2<j<1/2-m$ is fulfilled. In turn, the normalization condition
for $v_{j}$ has the form $-1/2-m<j<m+1/2$. Finally, there are five
normalized solutions for $u_{j}$ with $j=0,\pm 1,\pm 2$ when
$m=-5/2$, and one solution for $v_{j}$ with $j=0$ when $m=1/2$. In
the limit $\delta\rightarrow 0$ we arrive at the solution for
spherical fullerenes
\begin{eqnarray}
v_j(r) = r^{(-1/2-j-m)}(R^{2}+r^{2})^{(1/2+m)}e^{\frac{BR^{4}}{R^{2}+r^{2}}},\nonumber\\
u_j(r) =
r^{(-1/2+j+m)}(R^{2}+r^{2})^{(1/2-m)}e^{-\frac{BR^{4}}{R^{2}+r^{2}}}.\label{equv}
\label{eq:5.15}
\end{eqnarray}
Since normalization conditions do not depend on the parameter
$\delta$ they are the same as for spherical case. Thus the
perturbation does not change the number of zero modes. Notice that
Eq. (\ref{eq:5.15}) can be obtained from Eqs. (\ref{eq:5.13}) and
(\ref{eq:5.14}) by using $\lim\left[1+(x/a)\right] ^{a}=e^{x},
a\rightarrow\infty$. Because of the different behaviour of the
zero-energy modes there is also possibility to recognize the
zero-eigenvalue states in experiment.

\section[]{The $SU(2)$ algebra}
The angular-momentum operators for Dirac operator on the sphere
$S^2$ with charge $G$ and total magnetic monopole $m$ looks as
follows
\begin{equation}
\hat{L}_{z}=-i(\partial_{\phi}\mp i G),\label{eq:5.56}
\end{equation}
\begin{equation}
\hat{L}_{x}=i\sin\phi\partial_{\theta}+i\cos\phi
\frac{\cos\theta}{\sin\theta}\left(\partial_{\phi}\mp
iG-i\frac{m}{\cos\theta}\right)+\sigma_{z}\frac{\cos\phi}{2\sin\theta},\label{eq:5.57}
\end{equation}
\begin{equation}
\hat{L}_{y}=-i\cos\phi\partial_{\theta}+i\sin\phi
\frac{\cos\theta}{\sin\theta}\left(\partial_{\phi}\mp
iG-i\frac{m}{\cos\theta}\right)+\sigma_{z}\frac{\sin\phi}{2\sin\theta},\label{eq:5.58}
\end{equation}
where $-(+)$ sign correspond to north (south) hemisphere. These
operators satisfy the standard commutations relations of the $SU(2)$
algebra:
\begin{equation}
\varepsilon_{ijk}\hat{L}_{j}\hat{L}_{k}=i\hat{L}_{i}.\label{eq:5.59}
\end{equation}
The square of Dirac operator and $\hat{L}^{2}$ may be diagonalized
simultaneously
\begin{equation}
\hat{\cal{D}}_{0}^2(B=0)=\hat{L}^{2}+\frac{1}{4}-m^{2}.\label{eq:5.60}
\end{equation}
We can see that the operator $V(\theta,\phi)$ in the case when the
magnetic field is pointed in the z direction can be expressed in the
form
\begin{equation}
V(\theta,\phi)=\hat{L_{z}}+\left(\frac{\sigma_{3}}{2}-m\right)z.\label{eq:5.61}
\end{equation}
In the case when the magnetic field is pointed in the x direction we
have
\begin{equation}
V(\theta,\phi)=-\hat{L_{x}}-\left(\frac{\sigma_{3}}{2}-m\right)x,\label{eq:5.62}
\end{equation}
where x,z are Cartesian coordinate
\begin{equation}
x=\sin\theta \cos\phi; \quad  z=\cos\theta.\label{eq:5.63}
\end{equation}
The square of Dirac operator and operator $\hat{L}^2$ may be
diagonalized simultaneously and their eigenvalues are interrelated
\begin{equation}
\langle E,n| \hat{L}^{2} |E,n \rangle
=l(l+1)=E^{2}-\frac{1}{4}=(n+|j|)(n+|j|+1),
\end{equation}
where
\begin{equation}
l=n+|j|.
\end{equation}
We can use instead of $n$ the angular momentum $l=n+|j|$. We can
introduce the eigensteite of $\hat{\cal{D}}_{0}(B=0)$ in the form
$$
\Psi_{l,j}(x,\phi)=\frac{e^{i(j\pm G)\phi}}{\sqrt{2\pi}\Omega}
\sqrt{\frac{(l+j)!}{(l-j)!}}
$$
\begin{equation}
\times \left( \begin{array}{c}(1-x)^{-
\frac{1}{2}(j-m-1/2)}(1+x)^{-
\frac{1}{2}(j+m+1/2)}\frac{d^{l-j}}{dx^{l-j}}(1-x)^{(l-m-1/2)}(1+x)^{(l+m+1/2)} \\
i sgn(E)(1-x)^{-\frac{1}{2}(j-m+1/2)}(1+x)^{-
\frac{1}{2}(j+m-1/2)}\frac{d^{l-j}}{dx^{l-j}}(1-x)^{(l-m+1/2)}(1+x)^{(l+m-1/2)}
\end{array} \right)
\end{equation}
where
\begin{equation}
\Omega=2^{l}\sqrt{\Gamma(l-m+1/2)^{2}+\Gamma(l+m+1/2)^{2}},
\end{equation}
and $\Gamma$ are the Gamma functions (see Ref.~\cite{Abrikosov}).
When we act by $L_{-}$ and $L_{+}$ on $\Psi$ we find that
\begin{equation}
\hat{L}_{-}\Psi_{l,j}=\sqrt{(l+j)(l-j+1)}\Psi_{l,j-1},
\end{equation}
\begin{equation}
\hat{L}_{+}\Psi_{l,j}=\sqrt{(l+j+1)(l-j)}\Psi_{l,j+1},
\end{equation}
where
\begin{equation}
\hat{L}_{-}=-e^{-i\phi}\left(\partial_{\theta}-i\frac{\cos\theta}{\sin\theta}\left(\partial_{\phi}\mp
iG-i\frac{m}{\cos\theta}\right)-\frac{\sigma_{z}}{2\sin\theta}\right),
\end{equation}
\begin{equation}
\hat{L}_{+}=e^{i\phi}\left(\partial_{\theta}+i\frac{\cos\theta}{\sin\theta}\left(\partial_{\phi}\mp
iG-i\frac{m}{\cos\theta}\right)+\frac{\sigma_{z}}{2\sin\theta}\right).
\end{equation}
Now we would like to transform our formulae to the cartesian
coordinates. The corresponding transformation rules for spinors
are~\cite{Abrikosov}
\begin{equation}
(\Psi)_{C}=V^{\dag} \Psi,
\end{equation}
and cartesian realization of operator $\hat{L}$ is
\begin{equation}
\hat{L}_{C}=V^{\dag} \hat{L}V,
\end{equation}
with $ V $-matrices given by
\begin{equation}
V=\exp\left(\frac{i\sigma_{y}}{2}\theta
\right)\exp\left(\frac{i\sigma_{z}}{2}\phi \right).
\end{equation}
For example we present the Cartesian realization of $\hat{L}_{x}$
operator
\begin{equation}
(\hat{L}_{x})_{C}=i\sin\phi\partial_{\theta}+i\cos\phi
\frac{\cos\theta}{\sin\theta}\left(\partial_{\phi}\mp
iG-i\frac{m}{\cos\theta}\right)+\frac{\sigma_{x}}{2},
\end{equation}
that may be expressed in the form ($m=G=0$)
\begin{equation}
(\hat{L}_{x})_{C}=y \hat{p}_{z}-z \hat{p}_{y}+\frac{\sigma_{x}}{2}.
\end{equation}
\newpage

\chapter[Electronic Structure of
Spheroidal Fullerenes]{Conclusion}

Electronic properties of fullerenes molecules have been discussed
from a theoretical point of view. The effective-mass approximation
is particularly suitable for understanding global and essential
features. In this scheme, the motion of electrons in fullerenes is
described by Dirac-Weyl's equation. The geometry and topology is
found to influence the main physical characteristics of graphite
nanoparticles, first of all, their electronic properties. The
topological defects (disclinations) appear as generic defects in
closed carbon structures.

We have considered the electronic states of spheroidal fullerenes
provided the spheroidal deformation from the sphere is small
enough. In this case, the spherical representation is used for
describing the eigenstates of the Dirac equation, with the slight
asphericity considered as a perturbation. The using of the
perturbation scheme allows us to find the exact analytical
solution of the problem. In particular, the energy spectrum of
spheroidal fullerenes is found to possess the fine structure in
comparison with the case of the spherical fullerenes. We have
shown that this structure is weakly pronounced and entirely
dictated by the topological defects, that is it has a topological
origin. We found three twofold degenerate modes near the Fermi
level with one of them being the true zero mode therefore our
finding confirms the results of~\cite{Saito} that there can exist
only up to twofold degenerate states in the $C_{70}$.

Notice that for spherical fullerenes ($\delta=0$) our results agree
with those found in~\cite{Osipov}. It is interesting that the
predictions of the continuum model for spherical fullerenes are in
qualitative agreement with tight-binding
calculations~\cite{Tang,Manousakis,Perez,Lin}. In particular, the
energy gap between the highest-occupied and lowest-unoccupied energy
levels becomes more narrow as the size of fullerenes becomes larger.
It is important to keep in mind, however, that the continuum model
itself is correct for the description of the low-lying electronic
states. In addition, the validity of the effective field
approximation for the description of big fullerenes is also not
clear yet. Actually, this approximation allows us to take into
account the isotropic part of long-range defect fields. For bigger
fullerenes, one has to consider the anisotropic part of the
long-range fields, the influence of the short-range fields due to
single disclinations as well as the multiple-shell structure.

We have focused also on the structure of low energy electronic
states of spheroidal fullerenes in the weak uniform magnetic
field. For the states at the Fermi level, we found an exact
solution for the wave functions. It is shown that the external
magnetic field modifies the density of electronic states and does
not change the number of zero modes. For non-zero energy modes,
electronic states near the Fermi energy of spheroidal fullerene
are found to be splitted in the presence of a weak uniform
magnetic field. The case of the x-directed magnetic field was
considered and compared with the case of the z-th direction. The z
axis is defined as the rotational axis of the spheroid with
maximal symmetry. The most important finding is that the splitting
of the electronic levels depends on the direction of the magnetic
field. Our consideration was based on the using of the
eigenfunctions of the Dirac operator on the spheroid, which are
also the eigenfunction of $\hat{L}_{z}$. Let us discuss this
important point in more detail.

In the case of a sphere there is no preferable direction in the
absence of the magnetic field. The magnetic field sets a vector,
so that the z-axis can be oriented along the field. In this case,
one has to use such eigenfunctions of the Dirac operator which are
also the eigenfunctions of $\hat{L}_{z}$. For the x-directed
magnetic field, the eigenfunction of both the Dirac operator and
$\hat{L}_{x}$ must be used. Evidently, the same results will be
obtained in both cases.

The situation differs markedly for a spheroid. The spheroidal
symmetry itself assumes the preferential direction which can be
chosen as the z-axis. In other words, the external magnetic field
does not define the preferable orientation. The symmetry is already
broken and, as a result, the case of the magnetic field pointed in
the $x$ direction differs from the case of the $z$-directed field.
For instance, there are no eigenfunctions which would be
simultaneously the eigenfunctions of both the Dirac operator on the
spheroid and $\hat{L}_{x}$. For this reason, the structure of the
electronic levels is found to crucially depend on the direction of
the external magnetic field (Figures~5.3-5.8).

It should be mentioned that the zero-energy states in our model
corresponds to the HOMO (highest occupied molecular orbital)
states in the calculations based on the local-density
approximation in the density-functional theory (see,
e.g.~\cite{Saito}). In particular, the HOMO-LUMO energy gap is
found to be about $1.1$ eV for YO-C$_{240}$ fullerene within our
model (here LUMO means the lowest unoccupied molecular
orbital)~\cite{Pudlak1,Pudlak2}.

The very big fullerenes like C$_{960}$ and C$_{1500}$ become more
deformed, faceted and can no longer form a free-electron model like
the electronic shell~\cite{miz}, which was the assumption for this
model. For these structures the deviation from the sphericity is
larger when the pentagon defects are localized at the opposite
poles. In the case when the poles are far away from each other we
obtain the structure of nanotubes, and for the exact description
some new model related to that proposed here should be used. In our
opinion, our predictions are quite general for the fullerene family
and are of interest for experimental studies. Finally, we think that
the spheroidal geometry approach especially our established generic
models could also be related to other physical problems with
slightly deformed spherical structures that are common in the nature
e.g. in astrophysics for modeling pulsars or kvasars.

\newpage \leftline{\bf Acknowledgment} \vspace{5mm}
We would like to thank especially Prof. V. A. Osipov whose is also
co-author our results presented in this review for usefully comments
and advices during writing this chapter. The work was supported in
part by VEGA grant 2/7056/27. of the Slovak Academy of Sciences, by
the Science and Technology Assistance Agency under contract No.
APVT-51-027904.

\vspace{30mm}

\begin{thebibliography}{99}

\bibitem{rajaraman}  R. Rajaraman: \textit {Solitons and
                     Instantons}, (North-Holland, Amsterdam 1982)
\bibitem{Volod} V.A. Osipov: Phys. Lett. A \textbf {164} (1992) 327
\bibitem{kroto} H.W. Kroto, J.R. Heath, S.C. O'Brien {\it et al.}: Nature \textbf {318}
(1985) 162
\bibitem{kratschmer} W. Kratschmer, L.D. Lamb, K. Fostiropoulos {\it et al.}: Nature \textbf {347}
(1990) 354
\bibitem{iijima} S. Iijima: Nature \textbf {354} (1991) 56
\bibitem{ebbesen} T.W. Ebbesen: Physics Today, (June) (1996) 26
\bibitem{Iijima1}S. Iijima, T.Ichihashi, Y.Ando: Nature \textbf {356}
(1992) 776
\bibitem{Dress} R.Saito, G.Dresselhaus, M.S.Dresselhaus: Phys. Rev. B \textbf {53}
(1996) 2044
\bibitem{terrones} H. Terrones, M. Terrones: Carbon \textbf{36}
(1998) 725
\bibitem{mele} D.P. DiVincenzo, E.J. Mele: Phys. Rev. B \textbf {29}
(1984) 1685
\bibitem{Wallace}P.R. Wallace, Phys. Rev. {\bf71} (1947) 622
\bibitem{Weiss}J.C. Slonczewski and P.R. Weiss, Phys. Rev. {\bf109} 272
(1958) 272
\bibitem{jetplet00} V.A. Osipov, E.A. Kochetov: JETP Lett. \textbf {72}
(2000) 199
\bibitem{jose92} J. Gonz\'{a}lez, F. Guinea, M.A.H. Vozmediano: Nucl.Phys. B \textbf {406}
(1993) 771
\bibitem{kane} C.L. Kane, E.J. Mele: Phys. Rev. Lett. \textbf {78}
(1997) 1932
\bibitem{jose91} J. Gonz\'{a}lez, F. Guinea, M.A.H. Vozmediano: Phys. Rev. Lett. \textbf {69}
(1992) 172
\bibitem{ando1} T.Ando: J.Phys.Soc. Jpn.\textbf {74} (2005) 777
\bibitem{hou} J.G. Hou, J. Yang, H. Wang {\it et al.}: Phys. Rev. Lett.  \textbf {83}
(1999) 3001
\bibitem{carroll} D.L. Carroll, P. Redlich, P.M. Ajayan {\it et al.}: Phys. Rev. Lett.  \textbf {78}
(1997) 2811
\bibitem{kim} P. Kim, T.W. Odom, J.-L. Huang {\it et al.}: Phys. Rev. Lett.  \textbf {82}
(1999) 1225
\bibitem{an} B. An, S. Fukuyama, K. Yokogawa {\it et al.}: Appl. Phys. Lett. \textbf {78}
(2001) 3696
\bibitem{berber} S. Berber, Y.-K. Kwon, D. Tom\'anek: Phys. Rev. B \textbf {62}
(2000) R2291
\bibitem{jetp03} V.A. Osipov, E.A. Kochetov, M. Pudlak: JETP \textbf {96}
(2003) 140
\bibitem{kobayashi} K. Kobayashi: Phys. Rev. B \textbf {61} (2000) 8496
\bibitem{charlier} J.-C. Charlier, G.-M. Rignanese: Phys. Rev. Lett. \textbf {86}
(2001) 5970
\bibitem{yaguchi} T. Yaguchi, T. Ando: J.Phys.Soc.Jpn. \textbf {71}
(2002) 2224
\bibitem{jackiw77} R. Jackiw, C. Rebbi:   Phys. Rev. D \textbf {16}
(1977) 1052
\bibitem{jackiw81} R. Jackiw, J.R. Schrieffer: Nucl. Phys. B \textbf {190}
(1981) 253
\bibitem{salomaa} M.M. Salomaa, G.E. Volovik: Rev. Mod. Phys. \textbf {59}
(1987) 533
\bibitem{jackiw84} R. Jackiw: Phys. Rev. D \textbf {29} (1984) 2375
\bibitem{volovik99b} G.E. Volovik: JETP Lett. \textbf {70} (1999) 792
\bibitem{lutt} J.M.Luttinger, Phys.Rev.{\bf84} (1951) 814
\bibitem{lammert} P.E. Lammert, V.H. Crespi: Phys. Rev. Lett. \textbf {85}
(2000) 5190
\bibitem{jetplet01} V.A. Osipov, E.A. Kochetov: JETP Lett. \textbf {73}
(2001) 631
\bibitem{Nakahara} M. Nakahara,
                 {\it Geometry, Topology and Physics}, (Institute of Physics Publishing Bristol 1998)
\bibitem{Davies} N.D. Birrell and P.C.W. Davies,
                 {\it Quantum Fields in Curved Space}, (Cambridge 1982)
\bibitem{jpa99} E.A. Kochetov, V.A. Osipov: J.Phys. A: Math.Gen. \textbf {32}
(1999) 1961
\bibitem{pincak} R. Pincak, V.A. Osipov: Phys. Lett. A \textbf {314}
(2003) 315
\bibitem{ovrut} B.A. Ovrut, S. Tomas: Phys.Rev. D \textbf {43}
(1991) 1314
\bibitem{kroto1} H. Kroto, Reviews of Modern Physics, {\bf 69}
(1997) 703
\bibitem{Ryan} R.T. Chancey {\it et al.}, Phys. Rev. A {\bf 67}
(2003) 043203
\bibitem{Saito} S. Saito and A. Oshiyama, Phys. Rev. B {\bf 44}
(1991) 11532
\bibitem{Lomer}W.H. Lomer, Proc. Roy. Soc. (London) {\bf A227}
(1955) 330
\bibitem{Luttinger} J.M. Luttinger and Kohn, Phys. Rev. {\bf97}
(1955) 869
\bibitem{Ando} H. Matsumura and T. Ando, J. Phys. Soc. Jpn. {\bf67}
(1998) 3542
\bibitem{Dresselhaus} R. Saito, G. Dresselhaus and M.S. Dresselhaus,
{\it Physical Properties of Carbon Nanotubes} (Imperial College
Press, London, 2003)
\bibitem{Crespi} P.E. Lammert and V.H. Crespi, Phys. Rev. B {\bf 69}
(2004) 035406
\bibitem{Tersoff} J. Tersoff, Phys. Rev. B {\bf 46} (1992) 15546
\bibitem{Osipov} D.V. Kolesnikov and V.A. Osipov, Eur.Phys.Journ. B {\bf 49}
(2006) 465
\bibitem{Pincak} R. Pincak, Phys. Lett. A {\bf 340} (2005) 267
\bibitem{Clemenger} K. Clemenger, Phys. Rev. B {\bf 32} (1985) 1359
\bibitem{Aoki} H. Aoki and H. Suezawa, Phys. Rev. A {\bf 46} (1992) R1163
\bibitem{Gockeler} M.G\"{o}ckeler and T.Sch\"{u}cker, \textit{Differential geometry, gauge theories,
                    and gravity} (Cambridge University Press 1989)
\bibitem{Yang} T.T.Wu and G.N.Yang, Nucl. Phys. B {\bf 107} (1976) 365
\bibitem{Wu} T.T.Wu and G.N.Yang, Phys. Rev. D {\bf 12} (1975) 3845
\bibitem{Landau} L.D. Landau and E.M. Lifshitz, {\it Quantum Mechanics} (Elsevier Science, Oxford, 2003)
\bibitem{yoshida} M. Yoshida and E. Osawa, Fullerene Sci. Tech. {\bf 1}
(1993) 55
\bibitem{Lu} J. P. Lu and W. Yang, Phys. Rev. B {\bf 49} (1994) 11421
\bibitem{Broglia} R.A. Broglia, G. Col\`{o}, G. Onida, H.E. Roman, {\it Solid State Physics of Finite Systems}, (Springer 2004)
\bibitem{Abrikosov} A.A. Abrikosov,jr., Int. Journ. of Mod. Phys. A {\bf 17}
(2002) 885
\bibitem{Tang} A.Ch. Tang and F.Q. Huang, Phys. Rev. B {\bf 51}
(1995) 13830
\bibitem{Manousakis} E. Manousakis, Phys. Rev. B {\bf 44} (1991) 10991
\bibitem{Perez} A. Perez-Garrido, F. Alhama and J.D. Catala,
Chem. Phys. {\bf 278} (2002) 71
\bibitem{Lin}Y.L. Lin and F. Nori, Phys. Rev. B {\bf 49} (1994) 5200
\bibitem{Pudlak1} M. Pudlak, R. Pincak and V.A. Osipov, Phys. Rev. B {\bf 74}
(2006) 235435
\bibitem{Pudlak2} M. Pudlak, R. Pincak and V.A. Osipov, Phys. Rev. A {\bf 75}
(2007) 025201
\bibitem{miz}N. Mizorogi {\it et al.}, Chem. Phys. Lett. {\bf 378}
(2003) 598
\end{thebibliography}
\end{document}